# Numerical Analysis of the Impact of Water Temperature Setpoint and Energy Strategies on Indoor Pool Performance

Younes Benakcha*[1,2], Matthieu Labat[1], Ion Hazyuk[2], Stéphane Ginestet[1]

[1] LMDC, INSA/UPS Génie Civil, 135 Avenue de Rangueil, 31400 Toulouse cedex 04 France.

[2] Univ Toulouse, IMT Albi, INSA Toulouse, ISAE-SUPAERO, CNRS, ICA, Toulouse, France

*Corresponding author: benakcha@insa-toulouse.fr

**Abstract**

Indoor swimming pools (ISPs) consume significant amounts of electrical and thermal energy to ensure the heating of water and air, ventilation, and maintaining adequate humidity levels. This is measured in $GWh$ per year for large installations, such as Olympic swimming pools (SPs). In this paper, the problem is initially addressed using a phenomenological approach at steady state of the air-water coupling, based on a real case study. The aim is to identify the key phenomena and the constraints that are the most sensitive, including those related to water and air quality management. A key action lever is found in evaporation, and more specifically, water temperature and the indoor dewpoint temperature, which act as its precursors. In a second step, two different strategies were tested to reduce energy consumption for water heating. It was determined that a strategy which incorporates night setback in conjunction with a precise restart time yields a maximum gain of 4%. The second strategy aims to enhance the energy recovery of thermal solar panels by enabling slight overheating of the pool. Its large volume provides effective energy storage, with estimated energy savings of up to 17% for a $1°C$ overheating. This strategy appears to be a viable option, as it is straightforward to implement. However, the impact of water overheating on the energy consumption of AHU still needs to be analyzed and managed.

**Nomenclature**

| Latin symbols | Description | Unit |
|---|---|---|
| $A$ | Area | $m^2$ |
| $a_1, a_2, a_3$ | Constants | — |
| $C_p$ | Specific heat capacity | $J.kg^{-1}.K^{-1}$ |
| $[CO_2]_t, [CO_2]_{t-1}$ | $CO_2$ levels at time $t$ and $t-1$ | ppm |
| $[CO_2]_{ext}$ | Outdoor $CO_2$ level. | ppm |
| $[Chlorine]_t$ | Chlorine concentration in water at time $t$ | $mg.l^{-1}$ |
| $[Chlorine]_0$ | Initial chlorine concentration in water | $mg.l^{-1}$ |
| $E_{occ}$ | Rate of evaporation when the pool is occupied per unit area | $kg.h^{-1}.m^{-2}$ |
| $E_o$ | Rate of evaporation when the pool is unoccupied per unit area | $kg.h^{-1}.m^{-2}$ |
| $E$ | $CO_2$ emission rate by a person | $l.min^{-1}.pers^{-1}$ |
| $F_{NCL_{3w}}$ | Mass transfer of $NCl_3$ from water to air | $mg.l^{-1}.cm.s^{-1}$ |
| $h$ | Enthalpy | $J.kg^{-1}.K^{-1}$ |
| $h_{conv}$ | Heat exchange coefficient by convection | $W.m^{-2}.K^{-1}$ |
| $I_n$ | Air infiltration rate | $vol.h^{-1}$ |
| $I$ | Solar radiation | $W.m^{-2}$ |

| Symbol | Description | Units |
|---|---|---|
| $k_1$ | The pseudo-first-order reaction rate constant | — |
| $k_2$ | The first-order degradation rate constant | — |
| $K$ | Constant | — |
| $L_v$ | Latent heat of vaporization of water (2500) | $kJ.kg^{-1}$ |
| $\dot{m}$ | Mass flow rate | $kg.h^{-1}$ |
| $\dot{m}_{H_2O}$ | Humidification/dehumidification rate in AHU | $kg.h^{-1}$ |
| $N^*$ | Number of occupants per unit pool area | $pers.m^{-2}$ |
| $N$ | Number of occupants | $pers$ |
| $[NCl_3]_t$ | Concentration of $NCl_3$ at instant t | $mg.l^{-1}$ |
| $n$ | Constant | — |
| p | Partial pressure | $Pa$ |
| $\dot{Q}$ | Power / heat rate | $W$ |
| $Q_V$ | Volumetric voluntary regulatory flow | $m^3.h^{-1}.pers^{-1}$ |
| $T$ | Temperature | °C |
| $t$ | Time | $s$ |
| $U$ | Coefficient of surface heat transmission | $W.m^{-2}.K^{-1}$ |
| $u$ | Air speed | $m.s^{-1}$ |
| $V$ | Volume | $m^3$ |

**Greek Symbols**

| Symbol | Description | Units |
|---|---|---|
| $\alpha$ | Transmission coefficient | — |
| $\Delta t$ | Time step | $h$ |
| $\Delta p$ | Partial pressure difference | $Pa$ |
| $\lambda$ | Ventilation rate + infiltration | $vol.h^{-1}$ |
| $v_{NCL_{3w}}$ | Mass transfer coefficient | $cm.s^{-1}$ |
| ε | Emissivity of the surface | — |
| ρ | Density | $kg.m^{-3}$ |
| σ | Stefan-Boltzmann constant $(5.67.10^{-8})$ | $W.m^{-2}.K^{-4}$ |
| τ | Air fraction | — |
| ω | Humidity ratio | $g_v.kg_{da}^{-1}$ |

**Subscripts**

| | |
|---|---|
| $a$ | Air of the SP hall |
| $aux$ | Auxiliary equipment (Solar collector, boiler, etc.) |
| $air$ | To designate the air heating demand |
| $a\_renew$ | Air renewal |
| $b$ | Bottom |
| $blow$ | Blown air |
| $boiler$ | Boiler, modeling the district's hot water |
| $cond$ | Conduction |
| $conv\_a\ /conv\_w$ | To designate convective energy loss/gain (in air or water) |
| $dp$ | Dew point |
| $da$ | Dry air |
| $dehu$ | Dehumidification |
| $ext$ | Exterior |
| $env$ | Envelope of the swimming hall |
| $evap\_w\ /\ evap\_a$ | To designate evaporation loss energy loss (in water) / gain (in air) |
| $fw$ | Fresh water supply |
| $fa$ | Fresh air |
| $glyc$ | Glycol water |
| $g$ | Ground |
| $hum$ | Humidification |
| $in$ | Indoor |
| $m$ | Mix |
| $met\_w/met\_a$ | To designate sensible and latent energy gains related to occupant activity in both air and water environments. |
| $out$ | Outdoor |
| $pool$ | Pool |
| $rad\_w$ | To designate the radiative energy gains in water due to interaction with walls |
| $sw$ | Saturated air at water surface |
| $set$ | Setpoint |
| $s$ | Sides |
| $sol$ | To designate the positive solar energy gain through windows |
| $Tot$ | Total |
| $UA$ | Usable area |
| $va/v$ | Vapor at air temperature / vaporization |
| $w$ | Water |
| $WS$ | Water surface |
| $wt\_fw$ | To designate Energy loss related to daily water renewal for maintaining water quality in the pool |

**Acronyms**

| | |
|---|---|
| AHU | Air Handling Unit |
| ISAE- SUPAERO | Institut Supérieur de l'Aéronautique et de l'Espace |
| ISP | Indoor swimming pool |

| OSP | Outdoor swimming pool |
| PID | Proportional Integral Derivative |
| PVT | Photovoltaic and thermal solar panels |
| SP | Swimming pool |

# 1 Introduction

## 1.1 Context

Buildings are responsible for 40% of the world's energy consumption (around 64 $PWh$, (Drgoňa et al., 2020)), which results in significant greenhouse gas emissions. Consequently, the building sector is now the largest consumer, surpassing the transportation and industry sectors (Yuan et al., 2021). According to a French government report, the building sector is responsible for 123 million tons of $CO_2$ emissions and 44% of France's energy consumption each year (Ministère de la Transition écologique, 2016). In January 2014, the European Commission unveiled the energy and climate framework. It aims to achieve carbon neutrality by 2050 and, by 2030, to reduce carbon emissions by 40%, increase energy efficiency by 27%, and raise the share of renewable energy to 27% compared to the reference year 1990 (European Commission, 2014).

Sports facilities represent approximately 8% of the total energy consumption of buildings in Europe (Yuce et al., 2014). ISPs are among the most energy-intensive sports facilities, with an average consumption of $5200\ kWh.m_{WS}^{-2}.year^{-1}$ for continental climatic conditions and around $4300\ kWh.m_{WS}^{-2}.year^{-1}$ for Mediterranean climates (Trianti-Stourna et al., 1998). France has an ageing aquatic infrastructure of around 4,000 public SPs, 60 % of which were built more than 30 years ago (COSTIC, ADEME, 2022). These pools account for 20% to 30% of a municipality's total building energy consumption, underscoring the need for renovation efforts.

For the same dimensions, an ISP consumes up to three times more compared to an outdoor swimming pool (OSP) (Trianti-Stourna et al., 1998) or office buildings (Zuccari et al., 2017). Figure 1 shows the comparison of pool energy consumption with other types of buildings. In this figure, statistics from two sources have been compared. ISPs consume approximately $410\ kWh.m_{UA}^{-2}.year^{-1}$ according to data from Trianti-Stourna et al., 1998, and between 280 and 1570 $kWh.m_{UA}^{-2}.year^{-1}$ according to (Kampel et al., 2016) for various case studies in Norway, Sweden, England, and Greece. Despite these variations in energy consumption, they still consume significantly more than other buildings and sports facilities, which are limited to $210\ kWh.m_{UA}^{-2}.year^{-1}$. Here, the consumption in pools is normalized to the usable surface area ($kWh.m_{UA}^{-2}.year^{-1}$) rather than just the surface of the pool water ($kWh.m_{WS}^{-2}.year^{-1}$), to make a fair benchmarking between ISPs and the other buildings.

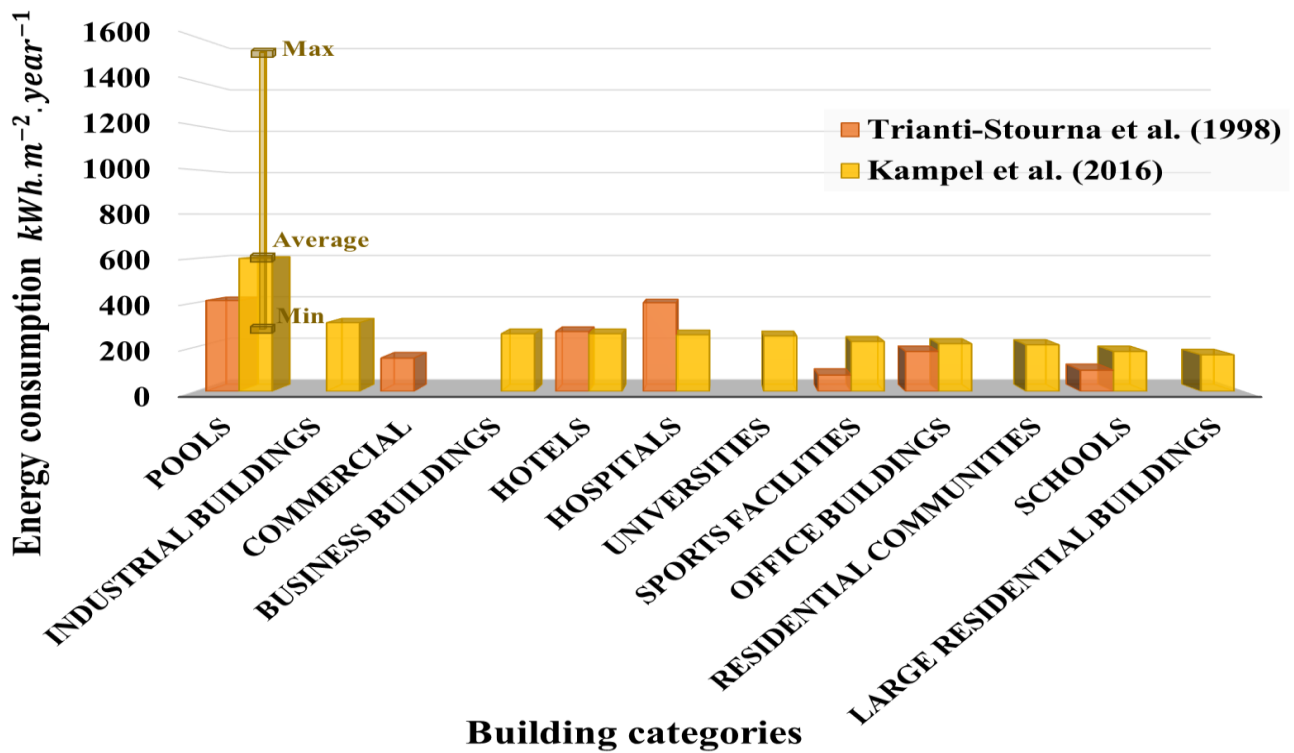

**Figure 1: Average annual energy consumption per usable area in several types of buildings.**

Energy sources can be categorized based on their end use into two main types: electrical energy, which powers ventilation, dehumidification systems, lighting, and pumps, and thermal energy, which is used for heating water (both for the pool and showers) and air heating (Delgado Marín and Garcia-Cascales, 2020). In Finland, the breakdown of energy usage of ISP is on average 66% of heat against 34% of electricity (Yuan et al., 2021). In the case study of the ISP in Kragujevac, Serbia, the authors collected electricity, gas, and water bills from 2014 to 2016 (Nikolic et al., 2021). It was found that the SP uses $691\ MWh$ of electricity, $261.10^3\ m^3$ of natural gas (equivalent to $2\ GWh$ of heat), and $37.10^3\ m^3$ of water every year. In (Yuan et al., 2021), a medium-sized SP (with an approximate usable surface area of 500 m²) consumes around 124 MWh of electricity and 208 MWh of heat per year, and its annual water consumption is around $2.10^3\ m^3$. The same document also provides information on the consumption levels for pools of other sizes. The significant energy consumption heterogeneity highlighted above illustrates the strong sensitivity of pool energy use to factors such as size, geographic location, and the type and number of HVAC (heating, ventilation, and air conditioning) systems. Each facility has a distinct configuration, as equipment varies widely between pools. This diversity results in substantial differences in energy consumption, making it difficult to define a standard reference pool, each is effectively unique. This is why the present research is based on a real case study in France to support the modeling work.

The energy consumption in the pool water circuit is mainly driven by two factors: the energy consumption of various pumps, particularly the filtration circuit pump, and pool water heating. Pumps energy consumption can reach up to 30% of the total electrical energy use (A. Saari and T. Sekki, 2008), while the energy required for water heating can account for a significant portion, up to 50% of the total energy consumption (Nikolic et al., 2021).

Water heating HVAC systems, such as boilers, are regularly engaged, especially during cold periods, to offset heat losses, meet energy demand, and ensure occupant comfort by maintaining the water temperature at the desired level. The significant energy demand for pool water heating is primarily attributed to evaporation occurring from the pool water into the surrounding air, which accounts for at least 60% of heat loss in the water ((Yuan et al., 2021), (Zuccari et al., 2017), (Delgado Marín and Garcia-Cascales, 2020), (Ribeiro et al., 2016)). In other words, evaporation not only causes cooling water, but it has also a direct impact on the humidity of the indoor air, which must be treated using AHUs, which are highly energy-intensive systems. Covering the water surface during unoccupied periods may be the simplest way to reduce evaporation, although it is not adopted in all SPs (Francey et al., 1980), (Czarnecki, 1963). The daily renewal of pool water, intended to maintain its quality, with fresh water from the municipal network, which must be reheated to the setpoint temperature, may also impose additional load on the HVAC systems. Moreover, leaks and conduction losses further amplify this load to meet the energy demand. To offset these heat losses, water heating can be provided by various types of equipment depending on the pool configuration: conventional thermal systems such as heat pumps and gas condensing boilers (Lam and Chan, 2001) or renewable energy installations such as hybrid solar systems, geothermal systems, and biomass boilers. Several scientific papers have been published which introduce advanced technologies for heating pool water using hybrid solar systems in combination with heat pumps, such as Solar Assisted Heat Pump (SAHP), Water Solar Assisted Heat Pump (W-SAHP), Photovoltaic/Thermal Solar Assisted Heat Pump (PVT-SAHP), and Direct Expansion Solar Assisted Heat Pump (DX-SAHP) (Tagliafico et al., 2012). These systems have been extensively studied, particularly for their potential to maximize the use of solar energy and enhance overall energy efficiency. By using a preheated source (thanks to solar energy), the heat pump operates under better thermodynamic conditions, increasing its coefficient of performance (COP), which also depends on the type of the technology involved. This helps reduce reliance on fossil fuels, leading to a lower environmental impact. The optimal surface area of the solar collector can be determined by analyzing the heat losses in the pool, particularly through evaporation and convection (Olcay et al., 2012). Other researchers went a step further by using a straightforward transient analytical method to derive an explicit formula for the water temperature in an ISP coupled to thermal collectors (Singh et al., 1989). Overall, it is safe to assume that replacing old, energy-consuming systems with solar-based devices would help to reduce the energy bill for ISP. However, the high cost and limited profitability of such systems in SPs may hinder their adoption (Buonomano et al., 2015). Even a conventional solar installation can quickly become unprofitable if it is oversized (Olcay et al., 2012). It can be profitable for small installations in warm climates (Lugo et al., 2019). Improving solar collector designs is of interest to some researchers. For instance, conventional flat plate solar collectors are cost-effective and easy to maintain but have low efficiency. In a study (Kumar and Kumar, 2024), conventional riser tubes were replaced with dual spiral-shaped flow tubes, leading to a 13.7% efficiency enhancement (up to 70.8%), and improved heat transfer by increasing surface contact and flow time. This is not the focus of the present study. One potential strategy for achieving energy savings is to implement energy harvesting solutions, such as the recovery of waste heat from shower water, a nearby ice rink chiller unit, or a data center. This can be accomplished by installing heat exchangers between the heat source and the pool water circuit, or, in the case of the ice rink, for instance, by storing the recovered heat in an underground thermal energy storage tank (Kuyumcu et al., 2016). Implementing these energy recovery systems also necessitates a greater investment. Finally, a third approach involves improving the control and optimal management of existing equipment, which can be a cost-effective and relevant solution for smaller budgets (Ribeiro et al., 2016), (Delgado Marín et al., 2019). This becomes particularly interesting when the pool is equipped only with a conventional solar thermal system that can benefit from the potential of free solar energy, and the challenge then lies in optimizing its use through effective management. For example, there are studies indicating the possibility of controlling only water temperature by implementing the method of Early Switch Off (ESO) on heating equipment (Delgado Marín et al., 2019; Fadzli Haniff et al., 2013). This involves switching off the boiler early at night when solar collectors will produce enough energy during the next day, based on weather predictions integrated into their control algorithm. According to (Delgado Marín et al., 2019), ESO predictive control reduces pool energy demand by 19% and fuel use by 43% compared to the use of a conventional PID control. These results are very promising, and to the best of the author's knowledge, no similar results were reported in the literature for other case studies. For this reason, it was decided to expand the search for potential energy savings through optimized control in the framework of water heating in ISP by maximizing the use of solar energy. The main contributions of this paper are, first, the identification, through a novel phenomenological air-water coupling analysis under steady-state conditions, that water temperature is a major driver of evaporation, leading to increased water heating demand. This is followed by a numerical evaluation, under dynamic conditions, of two strategies with different levels of complexity aimed at reducing water heating demand. The study concludes that the simplest strategy, allowing slight overheating from solar thermal energy, is significantly more advantageous than a strategy requiring more extensive modeling efforts. The latter was inspired by previous work of Delgado Marín et al., (2019) (methodological differences with the present study are detailed in Section 6) but yielded dissimilar results, as each pool represents a distinct case study.

## 1.2 Objectives of this study

In the present study, two energy management strategies comparable to the above-mentioned research, will be tested on a case study that is located in Toulouse, France. The first objective of this paper is to conduct a thorough analysis of the heat and mass transfer that occur in ISP, with a particular focus on those involved in the coupling between air and water circuits. This will be achieved under a steady-state assumption to discuss the impact of different settings on the energy demand and to identify leverage points, particularly for the water circuit. This involves a comprehensive understanding of the primary precursors of evaporation, a process that is responsible for the majority of heat losses in water.

The second goal is to implement a management strategy for the restart time of the heating equipment in the water circuit under dynamic simulation. The objective is to optimize the utilization of solar energy while simultaneously reducing reliance on thermal energy from an auxiliary source. This includes authorizing the overheating of pool water using solar energy, a strategy which can offer considerable advantages.

The key elements brought by this paper are:

1) Analyzing air-water coupled transfer in steady-state conditions and its impact on indoor energy consumption and air quality.

2) Assessing energy consumption, losses, and gains in the water circuit over a full year through dynamic simulation.

3) Numerically testing two strategies for enhancing solar energy use and reducing the use of auxiliary heating.

## 2 Physical phenomena

In order to ensure the proper functioning of air and water circuits, ISPs incorporate numerous HVAC systems. The complexity of managing these facilities lies in their simultaneous interaction with coupled phenomena, while each equipment's control remains independent. This section firstly presents the energy balances for the indoor air and the pool water volumes. Secondly, it focuses on the modelling of evaporation through a short literature review. Finally, indoor air quality issues will be depicted through the modelling of indoor concentration for $CO_2$ and trichloramines, and their removal by means of air renewal.

### 2.1 Energy balances

The pool's water can be idealized as a storage tank with energy inputs and outputs (see Figure 2). The energy balance on water is defined as the difference between the gains and losses of the various heat fluxes that influence water temperature. These fluxes can vary from one pool to another depending, among others, on the type and quantity of equipment used to heat the water, on the pool's size, its exposure to solar radiation, etc. Nevertheless, a general form can be adapted from several studies ((Buonomano et al., 2015), (Mancic et al., 2014), (Yuan et al., 2021)).

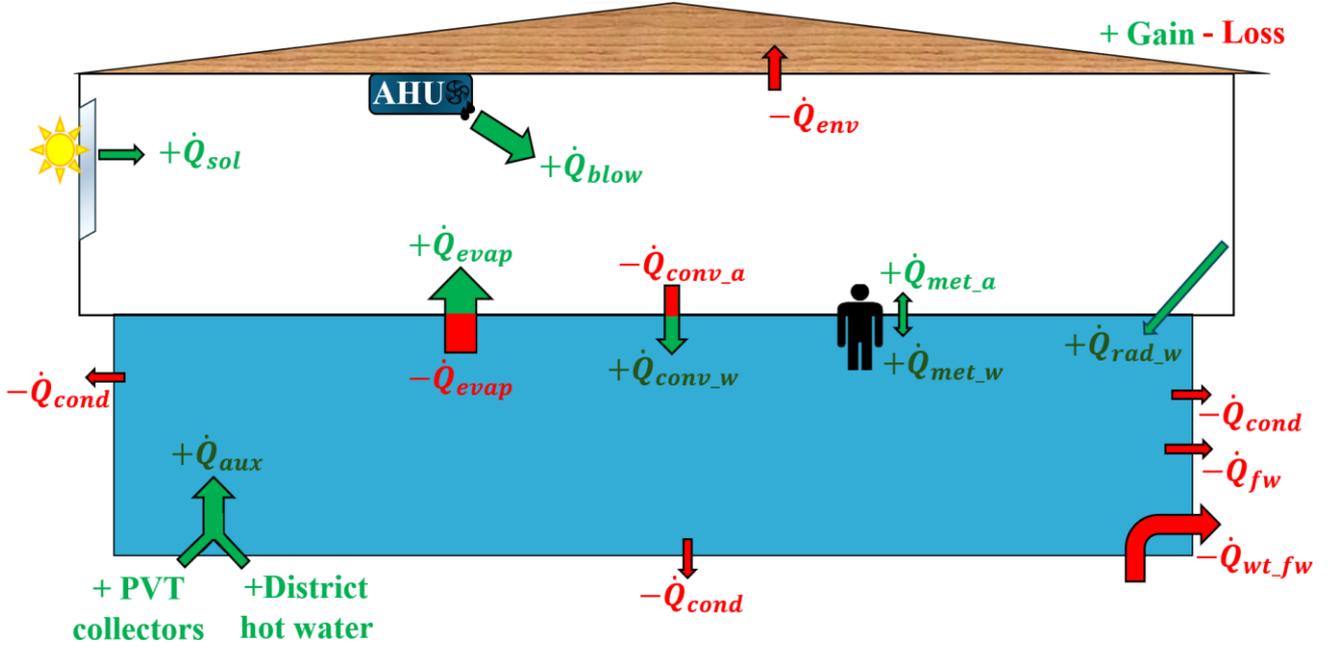

**Figure 2: Energy balances for air and pool water.**

The water balance is presented in equation (1) as follows:

$$\rho_w C p_w V_w \frac{dT_w}{dt} = \underbrace{\dot{Q}_{aux} + \dot{Q}_{met\_w}}_{Gains} + \underbrace{\dot{Q}_{conv\_w} + \dot{Q}_{rad\_w}}_{Gains\ or\ Losses} + \underbrace{\dot{Q}_{evap\_w} + \dot{Q}_{cond} + \dot{Q}_{fw} + \dot{Q}_{wt\_fw}}_{Losses} \tag{1}$$

The heat exchanges by convection (equation (2)) and radiation (equation (3)) are mainly driven by the temperature difference between water and the surroundings, respectively:

$$\dot{Q}_{conv\_w} = h_{conv} A_{pool} (T_a - T_w) \tag{2}$$

$$\dot{Q}_{rad\_w} = \sigma A_{pool} \varepsilon (T_{wall}^4 - T_w^4) \tag{3}$$

The conductive heat flux is related to the temperature difference between the pool and its direct environment, which is the ground. Conduction heat losses (equation (4)) can be described by considering the ground temperature at the vicinity, as well as the surface area of the bottom and the sides of the pool:

$$\dot{Q}_{cond} = U_{pool} (A_s + A_b)(T_g - T_{pool}) \tag{4}$$

Where $U_{pool}$ is the average conductive heat transfer coefficient of the pool's side and bottom surfaces. Some researchers consider that there are two distinct heat transfer coefficients and two different temperatures that distinguish the bottom ($U_b, T_{g,b}$) and the side ($U_s, T_{g,s}$) of the pool (Delgado Marín and Garcia-Cascales, 2020). This distinction was not considered in this paper for the sake of simplification.

Next, the pool is replenished with fresh water in order to compensate for evaporation and leakage ($\dot{m}_{fw}$, equation (5)). Also, a daily replacement of a proportion of the pool water ($\dot{m}_{wt\_fw}$, equation (6)) is mandatory in order to ensure its quality (Delgado Marín et al., 2019). Since the fresh water is supplied at ground temperature and must be heated to the temperature of the pool, two additional heat flows should be taken into account, as follows:

$$\dot{Q}_{fw} = \dot{m}_{fw}\, Cp_w\, (T_g - T_w) \tag{5}$$

$$\dot{Q}_{wt\_fw} = \dot{m}_{wt\_fw}\, Cp_w\, (T_g - T_w) \tag{6}$$

Swimmers generate sensible and latent heat because of their metabolism and physical activity ($\dot{Q}_{met\_w}$). This heat flux is frequently assumed to be marginal compared to others, and thus excluded from consideration. However, the metabolic heat gain can reach $525\,W$ per person during high-intensity activities such as swimming according to ASHRAE. Also, moisture production as high as $464\,g_v.h^{-1}.pers^{-1}$ can be expected. During periods of physical exertion, water vapor is emitted through perspiration and respiration. However, within the water environment, swimmers exhibit minimal perspiration, with the majority of the heat generated during muscle contractions being transferred directly to the water via convection. Consequently, the water vapor production of swimmers is predominantly attributed to respiration and is not expected to vary substantially compared to an individual engaged in routine activity ($185.2\,g_v.h^{-1}.pers^{-1}$). The difference between these two moisture flows multiplied by the latent heat of vaporization can be then considered as another heat gain in the water balance. Finally, an auxiliary heat gain ($\dot{Q}_{aux}$) is required to maintain the pool's temperature setpoint. The energy can be provided by a biomass/gas boiler, solar thermal panels, district heating or a geothermal installation, among other possibilities.

Next, the air energy balance is illustrated through the variation of enthalpy for indoor air (equation (7)). This enables the incorporation of evaporation into the global balance:

$$\rho_a V_a \frac{dh_a}{dt} = \underbrace{\dot{Q}_{blow} + \dot{Q}_{met\_a} + \dot{Q}_{sol} + \dot{Q}_{evap\_a}}_{Gains} + \underbrace{\dot{Q}_{conv\_a}}_{Gain\ or\ Loss} + \underbrace{\dot{Q}_{env}}_{Loss} \quad (7)$$

Indoor air is primarily heated by the components of an AHU, $\dot{Q}_{blow}$. Additionally, the pool can benefit from solar gains during the day, especially in summer. The solar gains through windows (Hazyuk et al., 2012) depends on transmission coefficient (α), window area ($A_{window}$) and solar radiation ($I$). It can be calculated using equation (8) as follows:

$$\dot{Q}_{sol} = \alpha\, A_{window}\, I \quad (8)$$

As for the water balance, the occupants contribute to heating and humidifying the indoor air ($\dot{Q}_{met\_a}$). The envelope's heat loss depends on the temperature difference between the interior and exterior (equation (9)):

$$\dot{Q}_{env} = U_{env}\, A_{env}(T_{ext} - T_a) \quad (9)$$

The convection and evaporation terms, $\dot{Q}_{conv\_a}$ and $\dot{Q}_{evap\_a}$, are of equal magnitude but of opposite signs to those in water balance equation ($\dot{Q}_{conv\_w} = -\dot{Q}_{conv\_a}$ and $\dot{Q}_{evap\_w} = -\dot{Q}_{evap\_a}$). For convection, the sign might change depending on the temperature difference between water and air. For evaporation however, it is always an energy loss in water and a gain in air.

Thus, the total energy balance of the ISP can be expressed by two distinct equations, one pertaining to air and the other to water. These two equations are coupled by convection and evaporation terms ($\dot{Q}_{conv\_a}$, $\dot{Q}_{conv\_w}$, $\dot{Q}_{evap\_a}$, and $\dot{Q}_{evap\_w}$). As previously stated, evaporation is the process that results in the loss of water mass, as well as leading to the cooling of the water due to latent heat (Trianti-Stourna et al., 1998). Given its substantial contribution to overall heat loss, if not its predominant role, the study of evaporation has been a subject of considerable interest over an extended period. The modeling of this phenomenon is explored in the next section.

## 2.2 Evaporation

Occupied pools present higher evaporation rates compared to unoccupied pools because the presence of swimmers in water increases water-air contact surface by generating sprays and waves. Furthermore, the wetting of deck and the wet bodies of the occupants increase this water-air contact, and consequently evaporation rate (Shah, 2012). The air-water interface can be represented as a diffuse interface, of very low thickness (the order of magnitude is a few micrometers). For an unoccupied SP with an undisturbed water surface, the extremely thin air-water interface quickly saturates because of the molecules' mobility at the contact level of the two fluids. When the air velocity at the level of this interface is zero, evaporation will take place only by molecular diffusion mechanism (slow process). In the opposite case, i.e. in the presence of an air movement at the interface level, the saturated air in the thin layer will be transported away and substituted by drier room air (Shah, 2014). The evaporation in this case will take place by two mechanisms: molecular diffusion (slow process) and mostly by air movement (fast process that accelerates evaporation). The air movement can be induced by forced convection resulting from building ventilation systems in ISPs or wind velocity in OSPs. It can also result from the buoyancy effect that takes place in natural convection. This latter phenomenon occurs when the ambient air gets saturated at the vicinity of the water surface, resulting in a decrease of its density. The saturated air becomes then lighter than the ambient air, moves upward and removes the evaporated water. The denser and drier ambient air that is above, faster it moves downward and replaces the saturated air (Shah, 2014), (Shah, 2012). Forced convection is the dominant process in outdoor pools while natural convection is the dominant process in indoor pools. Nevertheless, most researchers often consider forced convection to estimate evaporation rate in ISPs (Shah, 2014).

Dozens of correlations have been developed to estimate the evaporation rate in OSPs (Buonomano et al., 2015) and ISPs for both occupied and unoccupied scenarios. Since 1876, numerous researchers have proposed equations, as listed in (Shah, 2014), to calculate the evaporation rate for unoccupied pools. The following two forms (equations (10) and (11)) are the most common:

$$E_o = (a_1 + a_2 u)(p_{sw} - p_{va}) \qquad (10)$$

$$E_o = a_3 (p_{sw} - p_{va})^n \qquad (11)$$

It is remarkable that the difference between the partial pressure of saturated water vapor at water surface temperature ($p_{sw}$) and the partial pressure of the water vapor in the air ($p_{va}$) is the primary driving force behind evaporation.

Shah is among the researchers who have made significant contributions to the calculation of evaporation rate in occupied and unoccupied indoor pools. He has proposed several correlations with updates from 1981 to 2013, and some of his models have been considered as references by ASHRAE (Shah, 2014). According to (Shah, 2012), the rate of evaporation per unit area for an unoccupied pool ($E_0$) is given by equation (12) as follows:

$$E_0 = \max \left[ K(\rho_a - \rho_{sw})^{\frac{1}{3}} (\omega_{sw} - \omega_a),\ 0.00005(p_{sw} - p_{va}) \right] \qquad (12)$$

With: $\qquad K = 40$ when $\rho_{sw} - \rho_a \leq 0.02 \qquad / \qquad K = 35$ when $\rho_{sw} - \rho_a > 0.02$

Shah, 2012 has shown that the evaporation rate per unit area in an occupied pool ($E_{occ}$) depends on the number of occupants per unit area of the pool ($N^*$), the evaporation rate per unit area without occupancy ($E_o$), and the difference in density between hall air ($\rho_a$) and saturated air at water interface ($\rho_{sw}$). It is given by equation (13):

$$\frac{E_{occ}}{E_o} = 1.9 - 21(\rho_a - \rho_{sw}) + 5.3\ N^* \qquad (13)$$

When $N^* < 0.05$, a linear interpolation can be performed between $\frac{E_{occ}}{E_o}$ at $N^* = 0.05$ and $\frac{E_{occ}}{E_o} = 1$.

Note that $\rho_a - \rho_{sw}$ is considered zero in this formula when $\rho_a - \rho_{sw}$ is negative.

The heat loss involved in evaporation ($\dot{Q}_{evap\_w}$) introduces the latent heat of evaporation ($L_v$), taken at pool temperature. It is defined by equation (14):

$$\dot{Q}_{evap\_w} = A_{pool} L_v E_{occ} \tag{14}$$

**2.3 Indoor air quality in SPs**

Air quality within SPs is a significant consideration, influencing the need for air renewal and therefore the heat and moist balance of the air. For this reason, this section develops some considerations related to air quality.

According to numerous studies, when the concentration of $CO_2$ exceeds 1000 ppm in a room, effects such as headaches, fatigue, sensations of dry throat, eye and skin irritations, drowsiness and loss of attention are felt. In addition, over time, the human body, having a memory, gets used to the concentration of $CO_2$ and loses its sensations, which can represent a serious issue. Regarding the $CO_2$ concentration in the swimming hall, it can be expressed as follows (Bonato et al., 2020), (Labihi et al., 2024):

$$[CO_2]_t = [CO_2]_{t-1} e^{-\lambda \Delta t} + \left([CO_2]_{ext} + \frac{E.N}{\lambda.V}.6.10^4\right)\left(1 - e^{-\lambda \Delta t}\right) \tag{15}$$

With: $\quad \lambda = \frac{Q_V\, N}{V} + I_n$

For a given number of occupants, equation (15) can be used to determine the minimal airflow rate of fresh air ($Q_V$) per occupant under steady state conditions that satisfies the threshold value for indoor $CO_2$ concentration. It is given by the equation (16):

$$Q_V = \left(\frac{E.N}{V}.6.10^4 \frac{1}{[CO_2]_{Lim} - [CO_2]_{ext}} - I_n\right)\frac{V}{N} \tag{16}$$

Assuming that the building is pressurized, which can be achieved by applying supply airflow rate higher than the return airflow rate, the infiltration ($I_n$) can be neglected as in the present study, leading to the equation (17):

$$Q_V = 6.10^4 \frac{E}{[CO_2]_{Lim} - [CO_2]_{ext}} \tag{17}$$

The involvement of swimmers in a pool adds sweat, urine, saliva, hair, microorganisms, skin particles, and cosmetics to the water. These substances consist of organic carbon and nitrogen compounds such as ammonium ions, creatinine, acid amides, alpha-amino acids, and especially urea (Schmalz et al., 2011). Gaseous chlorine and sodium hypochlorite are typically used to disinfect water and manage its acidity. $NCl_3$ is one of the disinfection by-products resulting from the reaction of chlorine with the nitrogen compounds. It has been observed to cause irritation to mucous membranes, and as such, it is a primary focus of risk prevention organizations and labor codes, particularly with regard to the prolonged exposure of workers in pools. In France, for instance, a threshold value of $0.5\ mg.m^{-3}$ for the concentration of $NCL_3$ in ambient air is recommended.

The key phenomena involved in the production of $NCL_3$ are presented in (Schmalz et al., 2011) and briefly summarized here. The authors assert that urea is the most important precursor of $NCL_3$, as it reacts with chlorine in water as follows (equation (18)):

$$Urea + HOCl \xrightarrow{k_1} NCl_3 \xrightarrow{k_2} Products \qquad (18)$$

Where: $k_1$ is the pseudo-first-order reaction rate constant; $k_2$ is the first-order degradation rate constant.

The decomposition of chlorine (equation (19)) and the formation of $NCl_3$ (equation (20)) in water can be modeled as follows:

$$[Chlorine]_t = [Chlorine]_0 \times e^{-k_1 t} \qquad (19)$$

$$[NCl_3]_t = k_1/(k_2 - k_1) \times (e^{-k_1 t} - e^{-k_2 t}) \times [Chlorine]_0 \qquad (20)$$

$NCl_3$ is volatile and can be carried into the air with evaporation. The mass transfer of $NCl_3$ from water to air can also be calculated using the Deacon boundary layer model (equation (21)), which is a simplification, under certain conditions, of the Fick model:

$$F_{NCL_{3w}} = v_{NCL_{3w}} \times [NCl_3] \qquad (21)$$

Where the liquid mass transfer coefficient ($v_{NCL_{3w}}$) is equal to $0.6 \times 10^{-3} cm.s^{-1}$ for quiescent (unoccupied) pools, $2.4 \times 10^{-3} cm.s^{-1}$ for rippled (normal use) pools and $4.4 \times 10^{-3} cm.s^{-1}$ for rough surfaces (whirlpool).

Schmalz et al., 2011 indicate that ventilation and the state of the water surface are key factors in determining the concentration of $NCl_3$ in the air, whilst the impact of the chlorine concentration in the water is less direct. It has been demonstrated that the rate of transfer of $NCl_3$ from water to air is directly proportional to the degree of disturbance of the surface of the water. When the surface of the pool's water is undisturbed, the time for $NCl_3$ to be transferred into the air is prolonged. Consequently, the risk is minimal, as the turnover rate of the water treatment cycle (6 to 8 hours for a semi-Olympic pool) is faster than the mass transfer of $NCl_3$ to the air (20 hours to 5.8 days).

A direct consequence is that the instantaneous production of $NCl_3$ can be closely related to human activity rather than to the concentration in water, just as is the case for $CO_2$. This result is of particular interest given that the modelling of indoor $CO_2$ concentration is directly based on occupancy (equation (15)), and can be used to estimate the concentration of $NCl_3$. The assumption that the concentration of $CO_2$ and $NCl_3$ demonstrate analogous dynamic behaviors was confirmed by (Lester T. Lee et al., 2023) based on experimental observations. The authors assume that $CO_2$ can be used as a proxy for monitoring $NCl_3$ levels, as it is straightforward to measure using low-cost infrared gas sensors. This assumption is further substantiated by the findings of (Nitter and Hirsch Svendsen, 2020).

## 3 Case study

### 3.1 General description

As stated in the introduction, the stock for SPs is characterized by significant heterogeneity and complexity, which poses challenges when attempting to establish clear typologies. Consequently, it can be challenging to determine a typical SP that would be representative. For this reason, the present study opted for a case study approach, with the objective of deriving authentic and reliable values from which to conduct the analysis. The values presented in this section are the result of a number of different processes. Initially, on-site visits were conducted, and subsequently, interviews were undertaken with the energy manager of the SP. Thirdly, the as-built documentation was consulted. Finally, measurements available at the energy management system were recovered, and these were enhanced with additional on-site measurements that were carried out for the purpose of this study.

The studied SP is inspired by the one on the campus of the ISAE SUPAERO engineering school in Toulouse, France (see Figure 3 and Figure 4). The SP is semi-Olympic, primarily intended for the students at this engineering school, as well as for some swimming clubs. The indoor air volume is $4039\ m^3$, and the water surface area is $375\ m^2$ (15m $\times$ 25 m), with a water volume of $581\ m^3$. The SP benefited from the French government's 2021 recovery plan. These funds were used, among others, to thermally isolate the building envelope (door and window sealing) and acquire new high-performance equipment. The equivalent surface thermal transmittance coefficient of the entire pool envelope is estimated at $0.82\ W.m^2.K^{-1}$. A highly efficient air handling unit (AHU), a new filtration technology with diatomaceous earth filters, and a hybrid PVT solar installation were installed. The AHU, type DESHU XP+ 225 SYM, operates in a dual-flow mode, allowing for the implementation of an air recycling strategy that deals with two major effects: the regulation of the fresh air fraction and the dehumidification process. The nominal ventilation airflow rate in this AHU is $19500\ m^3.h^{-1}$.

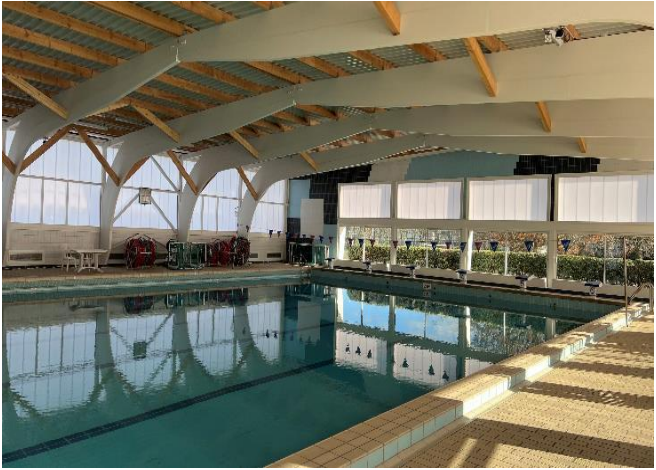 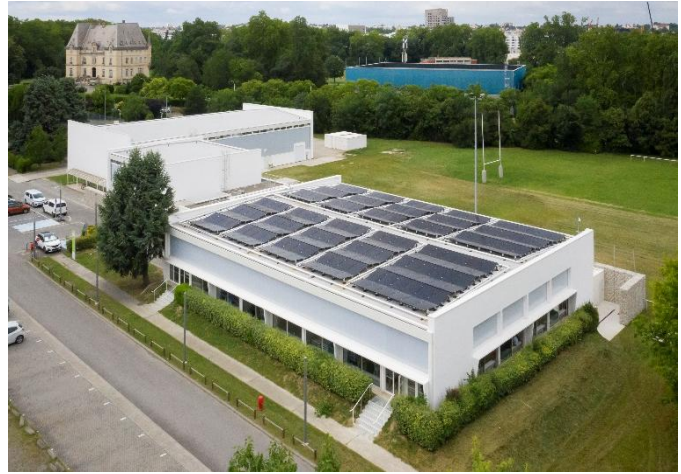

**Figure 3: Interior view of the ISAE SP**  **Figure 4: Exterior view of the ISAE SP**

In this paper, the focus is exclusively on the water circuit of the SP (see Figure 5), which consists of two circuits: the filtration circuit and the pool heating circuit:

- The filtration circuit comprises two parallel pumps, each with a maximum power of $30\ kW$, to ensure a maximum filtering flow of $290\ m^3.h^{-1}$. It should be noted that both pumps operate at approximately 53% of their maximum power, providing a total flow rate of $191\ m^3.h^{-1}$.

- The second circuit is dedicated to water heating, provided by two stages of heat recovery. This circuit is fed by a fraction of the water from the filtration circuit and operates continuously thanks to a pump with a flow rate of $38\ m^3.h^{-1}$ and a power of $1.5\ kW$. The water drawn by this pump passes through the first stage, where the heat is recovered by a heat exchanger connected to the solar circuit. This circuit includes hybrid solar collectors (PVT) with a total surface area of $495\ m^2$, a pump with a power of $632\ W$, and a flow rate of $18\ m^3.h^{-1}$. The second energy source is the district heating network, which transfers heat to the pool through the heat exchanger, with a capacity of 220 kW (see Figure 3).

The first heat recovery stage (PVT panels) is the primary source for water heating. It operates whenever conditions allow. If PVT heat production cannot maintain the pool temperature at the setpoint, the second heat source is activated. The various pumps and bypass valves are controlled locally using on/off logic with hysteresis, based on the temperature of the pool water and the temperature of the fluid at the outlet of the solar collectors. Should additional heat be required, a backup three-way valve combined with a PID controller will provide the necessary energy from a second source.

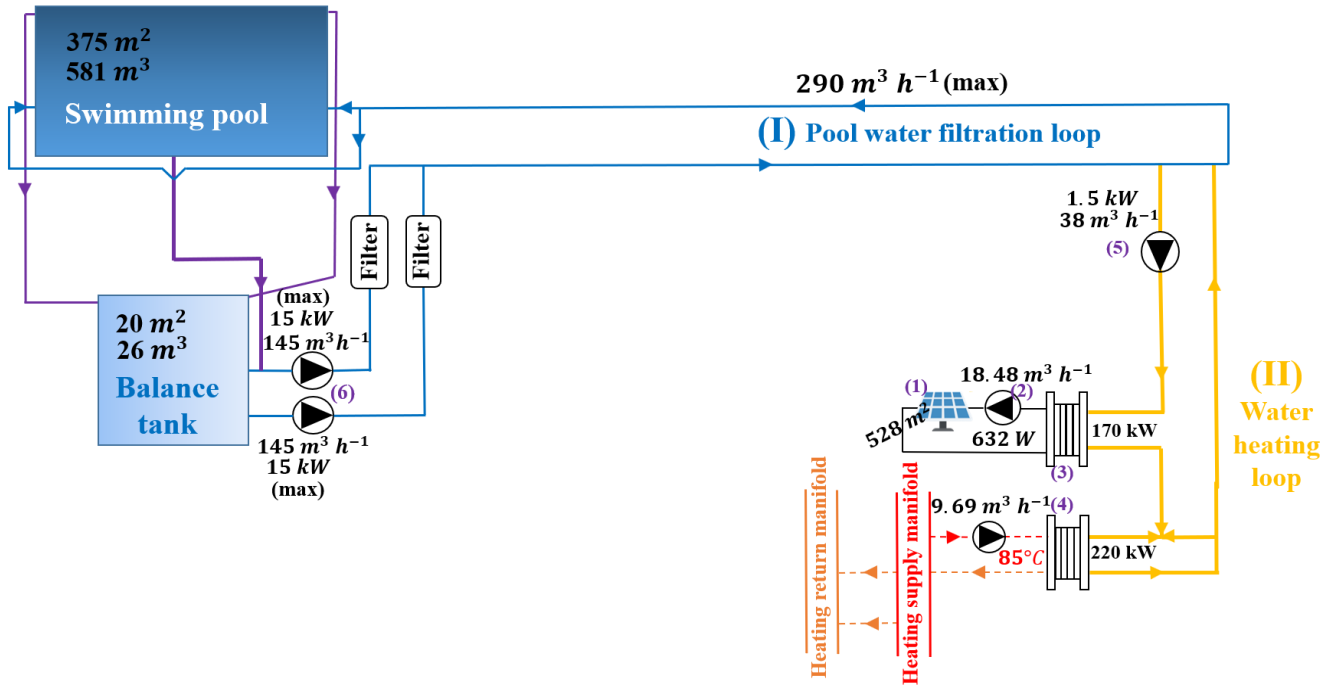

**Figure 5 : Diagram of the pool's water circuit. (1) PVT thermal collectors; (2) Solar circuit pump; (3) Solar circuit heat exchanger; (4) Heat exchanger with district heating network; (5) Pool heating circuit pump; (6) Filtration pumps.**

According to French legislation, a minimum daily renewal of 30 liters of non-recycled water per bather is required for public pools (MINISTÈRE DES SOLIDARITÉS ET DE LA SANTÉ, 2021). The daily use is estimated by the pool management operators to be between 150 and 200 people. This leads to approximately $6\ m^3$ of daily water renewal at ISAE pool. In other countries, a daily renewal of 5% of the pool volume is mandatory (Buonomano et al., 2015; Italian Standard UNI 10637:2006, 2006). At ISAE, the actual water renewal rate is $13\ m^3 \cdot day^{-1}$, including leak compensation. This is significantly higher than the minimum threshold set at the national level, but it represents only 2.24 % of the total water volume and is therefore below the recommendations for other countries.

### 3.2  Monitored values for indoor air and water

An on-site experimental campaign was conducted from 6 January 2025 to 6 February 2025. The campaign involved the utilization of five data loggers to monitor relative humidity and temperature at various locations. Two sensors were utilized for the monitoring of indoor air (KIMO type: KH-110 AO & KH-210 AO), one for outdoor air (type KH-110 AO), one for the air supplied by the AHU (KH-210 AO), and the last one for the air exhausted by the AHU (KH-200 AO). Prior to the commencement of the experimental campaign, a calibration process was undertaken to ensure the consistency of the measured values. For the purpose of temperature measurements, a PT100 sensor was utilized as a reference, with all sensors being placed within a climatic chamber. Three temperature setpoints were selected for the experiment ($15°C$, $25°C$, and $35°C$). Measurements were conducted at 15-minute intervals for a total test duration of 3 hours. With regard to the measurement of relative humidity, the five sensors were placed within a small container containing a saturated salt solution ($NaCl$ or $K_2CO_3$) in order to maintain a constant relative humidity level of 44% or 76%, respectively. These boxes were then placed in a climatic chamber to maintain a constant relative humidity of 50%. Each test was performed during at least 24 hours. The deviations observed for temperature measurements were consistent with the accuracy of the sensor employed as a reference. A slight difference in the monitored relative humidity was observed, which has been compensated in post-treatment for all sensors. Furthermore, the analysis of mass transfer is facilitated by the utilization of humidity ratio as opposed to relative humidity, due to the independence of humidity ratio from the temperature. Finally, water temperature was monitored daily by the pool's manager with a $\pm 0.5°C$ accurate sensor.

The measurements obtained during occupation are presented in Figure 6. However, it should be noted that measurements achieved during unoccupancy were discarded on account of the fact that an energy recovery strategy is applied for the AHU, which leads to significant temperature variations that complicate the analysis. It was considered that these variations would mainly affect the energy consumption of the AHU and that the impact on the water energy balance could be neglected as a first approximation.

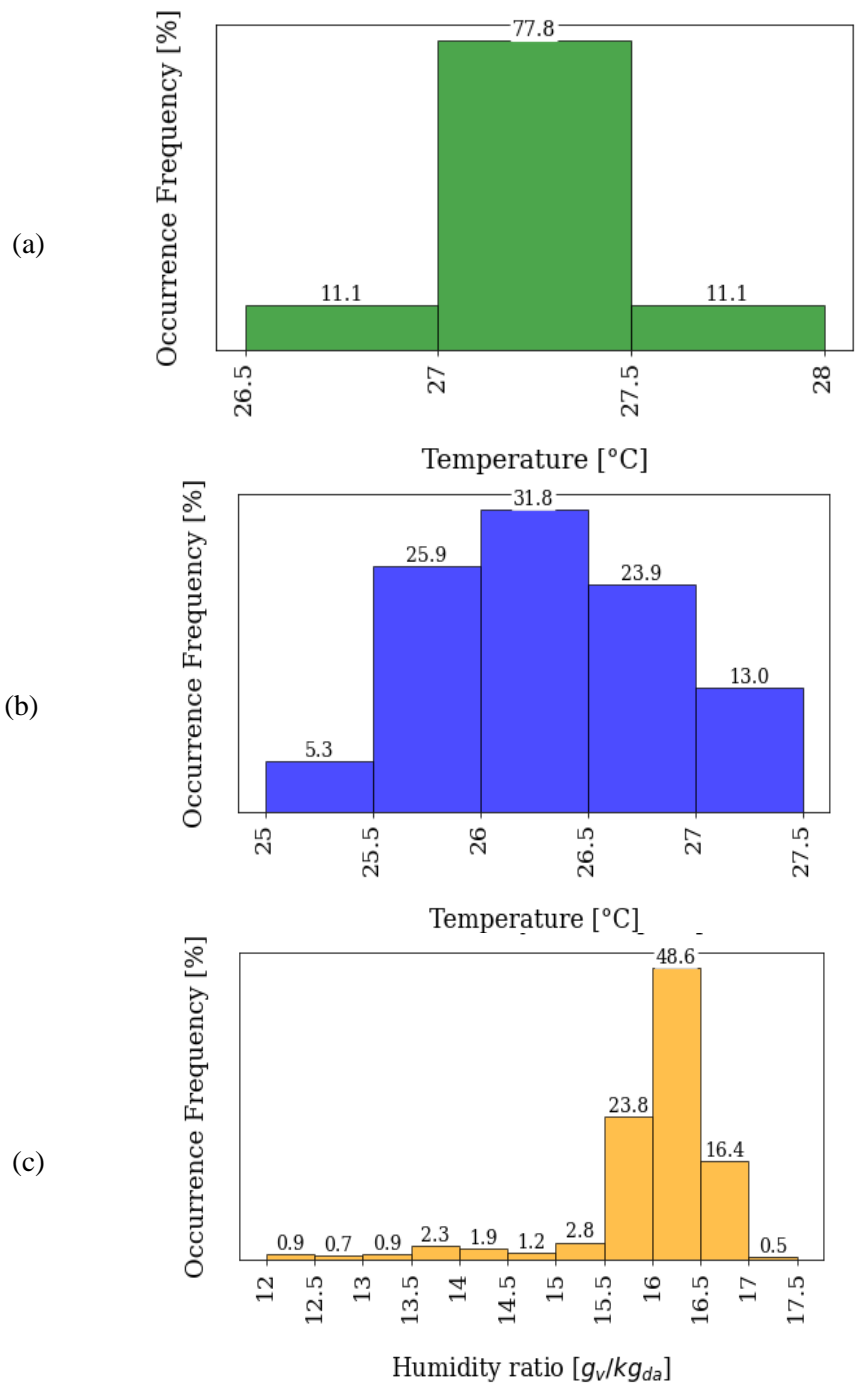

**Figure 6: Distribution of the monitored values during occupancy for water temperature (a), indoor air temperature (b) and indoor humidity ratio (c).**

First of all, it can be seen that the average water temperature remains within a narrow interval, and lies between 27 °C and 27.5 °C most of the time. This observation aligns with the temperature setpoint of 27 °C selected by the pool's manager, which remains constant throughout the year. Additionally, the use of a sensor with low accuracy underscores the fact that the measured values exceeding 27.5 °C are not meaningful. It should be noted that a temperature setpoint of 27 °C is within the recommended range but not commonly reported in the scientific literature for public SPs, where a value of 28 °C is more frequently observed. Although 28 °C is more comfortable for the swimmers, it has been found that the most accepted temperature range for competitions is between 26 and 28 °C (see Table 1). Moreover, a lower temperature setpoint would promote energy-saving strategies, a practice that was particularly stringent in Europe in recent years due to the rapid rise in energy costs. Furthermore, it is commensurate with the recommendations for sporting practice.

With regard to the monitored air temperature values, these vary between 25°C and 27.5 °C, with the most frequent values falling between 26 °C and 26.5 °C. These measurements are consistent with the temperature setpoint of 26 °C selected by the manager. The fairly large temperature variations are also consistent with the energy recovery strategy implemented on the AHU, which is not described in this paper for the sake of simplicity. These values are notably lower than the values found in the literature, which generally sets the temperature setpoint at $T_w$ + (1 to 2°C), typically fluctuating around 30°C (see Table 1). Another observation is that the air temperature setpoint is lower than for water, while the reverse is generally observed in the literature, as it is a common belief that it would increase the comfort of swimmers (Trianti-Stourna et al., 1998), (Ribeiro et al., 2016),(Yuan et al., 2021). Nevertheless, this is consistent with the ISAE SP strategy to minimize energy consumption.

**Table 1: Air and water temperature setpoints observed in scientific literature.**

|  | Water temperature ($T_w$) | Air temperature ($T_a$) |
|---|---|---|
| (Delgado Marín et al., 2019) | $25 - 28\ °C$ (International Swimming Federation).<br>$26°C$ (Spanish National Sports Council).<br>$28\ °C$ (Most of public SPs). | $T_w + 1°C$ |
| (Ribeiro et al., 2016) | $T_a - (1$ to $2\ °C)$ | $28 - 30\ °C$ |
| (Trianti-Stourna et al., 1998) | $26 - 28\ °C$ | $T_w + (1$ to $2\ °C)$ |
| (Yuan et al., 2021) | $26 - 28\ °C$ (Environmental Product Declaration).<br>$22 - 28\ °C$ (Other recommendations).<br>$28\ °C$ (Nowadays, due to increasing comfort demand). | $T_w + (1.5$ to $2.5\ °C)$ |

Finally, the indoor humidity ratio during occupancy varies between 12 and 17.5 $g_v.kg_{da}^{-1}$, with the most frequent values being between 16 and 16.5 $g_v.kg_{da}^{-1}$ which corresponds to the humidity setpoint. When combined with a temperature setpoint of 26 °C, this results in a targeted value of 75.2 % for indoor air relative humidity. It is also very common to see different values in the literature, such as 50 to 60 % in (Yuan et al., 2021). The promotion of such lower values is generally motivated by the objective of preventing the occurrence of condensation on the walls. However, given that the envelope of the ISAE building was retrofitted very recently, it can be stated that higher relative humidity can be undertaken with limited risk, which was confirmed by observations made on site. As illustrated in equation (12), increasing the indoor air humidity setpoint reduces the evaporation rate at the pool surface. Consequently, this results in a reduction of heat loss in water, leading to a decrease in energy demand. Therefore, increasing the setpoint for indoor humidity is a contributing factor to the energy saving strategy undertaken by the manager.

Based on these observations, the setpoints for water temperature, air temperature, and humidity ratio that will be considered as a reference in the continuation of the study are as follows: $T_w = 27\ °C$, $T_a = 26\ °C$ and $\omega_a = 16\ g_v.kg_{da}^{-1}$.

## 4 Analysis of air-water coupling and its impact on energy consumption

As illustrated in section 2, a strong coupling between pool water and indoor air exists and is influenced by numerous parameters, including the temperature and humidity setpoints selected by the pool manager. However, as emphasized in section 3.2, the observed values on site for the case study deviated from those commonly reported in the scientific literature, with the objective of promoting energy savings. Consequently, it was proposed to illustrate and analyze the influence of the setpoints on the air and water energy balances, as well as indoor air quality. This will be achieved in this section under the assumption of steady state conditions through the determination of:

1. The influence of water setpoint temperature on the evaporation rate;
2. The determination of the air renewal mass flow rate that satisfies indoor air quality requirements and its influence on the demand for dehumidification;
3. The energy demand under given steady-state conditions, with a decomposition of the demand for pool water heating and for the AHU.

### 4.1 Evaporation rate and water temperature setpoint

As mentioned earlier, the main phenomenon that causes interactions between pool water and hall air is evaporation. The effect of evaporation, on one hand, is pool water cooling, which impacts energy consumption and on the other hand it increases humidity of hall air which requires ventilation and dehumidification to avoid condensation on the walls of the building. The evaporation rate was calculated using the following method:

- o The partial pressure of saturated vapor ($p_{sw}$) at different values of $T_w$ is determined;

- The partial pressure of vapor ($p_{va}$) in indoor air at $T_a$ is set lower than $p_{sw}$, such that $p_{va} = p_{sw} - \Delta p$, where $\Delta p$ varies arbitrarily from 0 to 2500 Pa by step of 500 Pa. Then the dew point temperature ($T_{dp}$) is deduced from $p_{va}$;
- The remaining parameters such as relative humidity, humidity ratio, the density of saturated air ($\omega_{sw}$ and $\rho_{sw}$) at $T_w$ and of indoor air ($\omega_a$ and $\rho_a$) are determined through the means of established relationships in the field of air conditioning;
- The evaporation rate under occupied and unoccupied conditions can be finally computed from equations (12) and (13).

Figure 7 shows the evaporation rate ($\dot{m}_v$) as a function of the dew point temperature ($T_{dp}$) of the indoor air, for a constant air temperature ($T_a$) and different water temperatures ($T_w$). $\dot{m}_v$ is obtained by multiplying $E_0$ (equation (12) for unoccupied pool) or $E_{occ}$ (equation (13) for occupied pool) by the pool surface area.

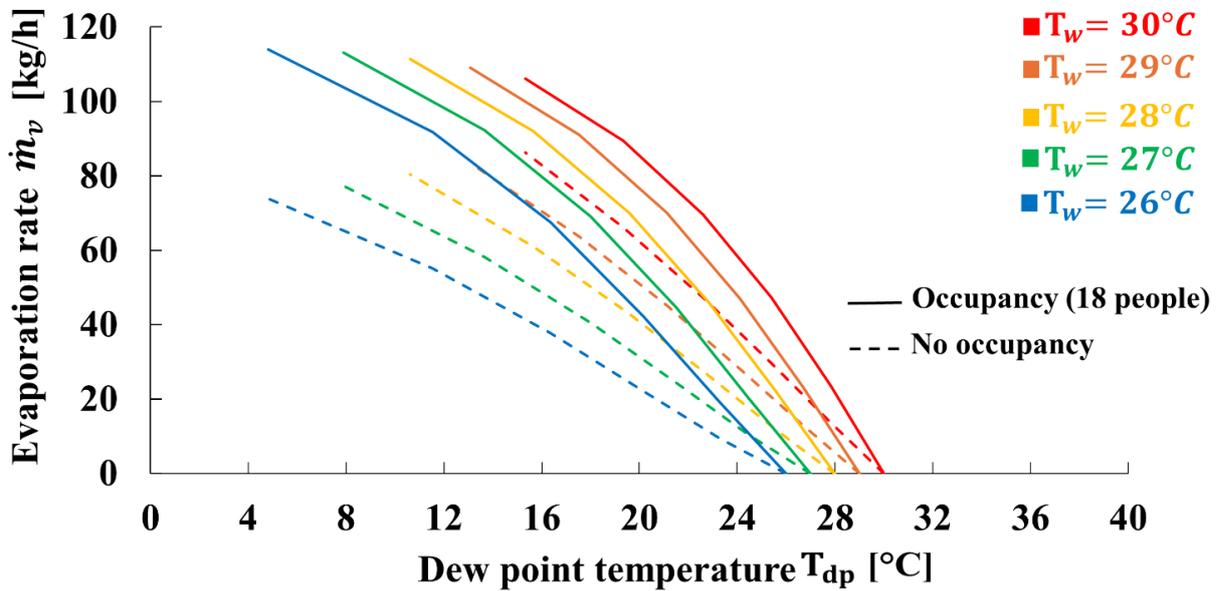

**Figure 7: Evaporation rate ($\dot{m}_v$) as a function of the dewpoint temperatures for different water temperature set points under a constant air temperature (26 °C).**

In the methodology proposed here, the dewpoint temperature is representative of the indoor humidity ratio, but also of the risk for condensation on the walls. If the wall surface temperature is lower than this value, condensation happens, which cannot be accepted. Therefore, to avoid the risk of condensation, a trade-off must be done between the insulation level of the walls, that increases the indoor temperature surface, and the indoor humidity ratio. In the ideal situation of perfectly thermally insulated walls, the dewpoint temperature would be the same as the water temperature setpoint, resulting in a moisture saturated air with no condensation on the walls but no evaporation neither. As this is impossible for obvious reasons, indoor humidity and the resulting dewpoint temperatures are lower, which leads to evaporation as evidenced in Figure 7. As expected, lower values of $T_{dp}$ increase the evaporation rate. However, the influence of the water temperature setpoint is very pronounced as evidenced here: the evaporation rate during occupation increases by approximately 98 % for a dewpoint temperature of 20°C when the water temperature increases by 4°C (from $26°C$ to $30°C$). Under a constant water temperature, significant variations in dewpoint temperature should be achieved to result in equivalent outcomes (from 20 to 12.7°C for $T_w = 26°C$). This underscores the key role of the water temperature setpoint on the evaporation rate. However, it is important to recall that decreasing water temperature also results in a reduction in occupant comfort. It is therefore necessary to balance the risk of condensation, the comfort of swimmers and the evaporation rate when managing the water temperature setpoint.

## 4.2 Air renewal and moisture removal

Next, air quality issues are depicted through the calculation of the minimal value of the mass flow rate for air renewal, which will be represented here by the fraction of fresh air introduced in the air circuit, and its consequence on the dehumidification demand.

Indoor air quality is determined by its concentration in $CO_2$ and $NCl_3$ and the minimum air renewal mass flow rate ($\dot{m}_{a,NCl_3}$, $\dot{m}_{a,CO_2}$) are calculated, at steady state. The value of $\dot{m}_{a,CO_2}$ was obtained using equation (16). Assuming maximum occupancy of 18 people and a $CO_2$ concentration limit of 1000 ppm, it leads to a minimum mass flow rate $\dot{m}_{a,CO_2}$ of 670 $kg.\square^{-1}$. The calculation of $\dot{m}_{a,NCl_3}$ requires several calculation steps inspired by *Table 2* in Schmalz et al., 2011. Assuming a high threshold for $NCl_3$ in the air of $0.5\ mg.m^{-3}$, a constant concentration of $0.08\ mg.l^{-1}$ and a rippled water surface leads to a value for $\dot{m}_{a,NCl_3}$ equal to $6253\ kg.\square^{-1}$. The air renewal rate is finally defined in equation (22) by selecting the highest value, as follows:

$$\tau_{fa,min} = \frac{\dot{m}_{a,min}}{\dot{m}_{a,Tot}} = \frac{\max(\dot{m}_{a,NCl_3}, \dot{m}_{a,CO_2})}{\dot{m}_{a,Tot}} = \frac{\dot{m}_{a,NCl_3}}{\dot{m}_{a,Tot}} \approx 27\ \% \qquad (22)$$

Where $\dot{m}_{a,Tot}$ is the nominal ventilation flow rate in the AHU, which is $23400\ kg.\square^{-1}$. It represents an air change in the pool of approximately $1.5\ vol.\square^{-1}$.

Generally, a minimum fresh air fraction of 30% of the airflow of the AHU is required according to (Schmalz et al., 2011), which is close to the calculated value (27%). According to the pool manager for this case study, the same rate is applied in the real AHU during occupancy periods.

Introducing fresh air directly affects the humidity ratio of the mixed air ($\omega_m$, equation (23)), as follows:

$$\omega_m = \tau_{fa}\ \omega_{ext} + (1 - \tau_{fa})\ \omega_a \qquad (23)$$

The introduction of fresh air contributes to the removal of moisture, thereby compensating a proportion of the moisture flow resulting from evaporation. However, the residual moisture must still be removed by means of dehumidification process. The mass flow demand for dehumidification is defined in equation (24) as follows:

$$\dot{m}_{H_2O} = \dot{m}_{a,Tot}(\omega_{blow} - \omega_m) \qquad (24)$$

Where $\omega_{blow}$ is the humidity ratio of the blown air. Neglecting the moisture provided by swimmers, $\omega_{blow}$ (see equation (25)) is obtained from the steady state balance of water content in indoor air:

$$\omega_{blow} = \omega_a - \frac{\dot{m}_v}{\dot{m}_{a,Tot}} \qquad (25)$$

These three equations can be combined so that it gives equation (26):

$$\dot{m}_{H_2O} = \dot{m}_{a,Tot} \, \tau_{fa} \, (\omega_a - \omega_{ext}) - \dot{m}_v \qquad (26)$$

It is worth underlining that a negative value for $\dot{m}_{H_2O}$ indicates dehumidification, which is generally acceptable given that the majority of dehumidification energy can be recovered. Conversely, a positive value indicates humidification, which is undesirable in the context of a SP where humidification already occurs.

The calculations have been performed for four distinct values of the outdoor air temperature and humidity ratio, listed in Table 2, which are representative of various climatic conditions during winter and spring in Toulouse, France. These values were used to plot Figure 8, Figure 9, Figure 10, and Figure 11.

It should be noted that, although the minimum air renewal was identified to be 27 % for the present case study, this was varied for the purposes of illustration, with the results presented in Figure 8.

**Table 2: External climatic conditions depending on the season of the year.**

|   | Dry winter | Mild winter | Mild spring | Humid spring |
|---|---|---|---|---|
| $T_{ext}$ [°C] | 5 | 15 | 20 | 25 |
| $\omega_{ext}$ [$g_v \cdot kg_{da}^{-1}$] | 2 | 7 | 9 | 14 |

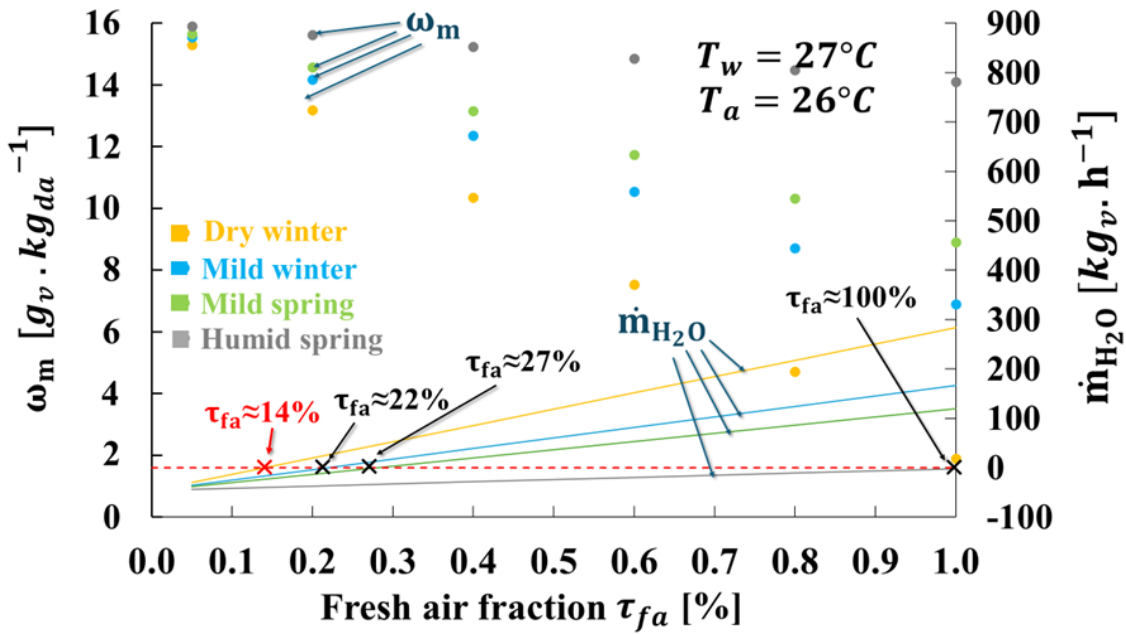

**Figure 8: Impact of fresh air fraction on the humidity ratio after mixing ($\omega_m$) and on the humidification/dehumidification demand ($\dot{m}_{H_2O}$) in AHU**

Figure 8 shows that drier outdoor conditions are associated with a reduction in the value of $\omega_m$. This is consistent with the assumption of a constant indoor moisture content. A significant decrease of dehumidification demand occurs when the fresh air fraction increases, and it can potentially transition into a humidification demand under most outdoor conditions. As previously mentioned, a humidification demand is undesirable and it can be concluded that a low fresh air fraction is preferable, especially in dry outdoor conditions, where the impact is most sensitive. Therefore, it can be assessed that the maximum fresh air fraction corresponds to the value for which there is no humidification or dehumidification demand ($\dot{m}_{H_2O} = 0$) when the outdoor conditions are critical, that is, for the driest air. This value is indicated in Figure 8 by a red cross and corresponds to 14% for this particular case study. It's worth noting that maximum fresh air fraction that can be used to avoid humidification in winter is greater than the minimum required for $CO_2$ evacuation ($\approx$ 3 %), but lower than that for $NCl_3$ removal according to equation (22). This means that $CO_2$ levels can be managed safely and that it is $NCl_3$ concentration that determines the minimum fresh air flow fraction to ensure both air quality and hygrometric considerations.

It can be deduced that the demand for humidification/dehumidification could be reduced by regulating the fresh air fraction, thereby enabling the evaporation rate to be managed by air renewal for the majority of the time. However, this approach does not align with the solution identified in the ISAE SP, wherein a constant fresh air fraction setpoint was preferred. Consequently, the present study will adopt the same assumption ($\tau_{fa} = 30$ %) for the remainder of the study. This value, as explained previously, is inspired by the literature and the current value applied in the ISAE SP during occupancy periods, which is also consistent with the value calculated using equation (22).

## 4.3 Energy demand under steady state conditions

This section analyzes the total energy demand under steady state conditions, distinguishing between water and air circuits.

First of all, the water energy balance presented in equation (1) is adapted (see equation (27)) in order to compute the energy required to maintain the water temperature under steady state conditions.

$$\dot{Q}_{aux} = -(\dot{Q}_{evap\_w} + \dot{Q}_{cond} + \dot{Q}_{fw} + \dot{Q}_{wt\_fw} + \dot{Q}_{conv\_w} + \dot{Q}_{rad\_w} + \dot{Q}_{met\_w}) \tag{27}$$

The computation is achieved by using the numerical values presented in the case study, with the assumption that $T_g$ is 12 °C as in (Buonomano et al., 2015), $h_{conv}$ is $5\ W.m^{-2}.K^{-1}$ and $U_{pool}$ is $0.5\ W.m^{-2}.K^{-1}$. The air energy balance presented in equation (7) is adapted in a similar manner. Neglecting the solar heat loads leads to equation (28):

$$\dot{Q}_{blow} = -(\dot{Q}_{env} + \dot{Q}_{conv\_a} + \dot{Q}_{evap\_a} + \dot{Q}_{met\_a}) \tag{28}$$

Where $\dot{Q}_{blow}$ represents the additional power that the AHU must supply in order to reach the setpoint. The sensible part is dominated by heat losses through the envelope ($\dot{Q}_{env}$) while the latent heat is dominated by evaporation at the water's surface.

As a result, the energy demand for the AHU is related to three needs at steady state: dehumidification because of indoor moisture sources ($\dot{Q}_{dehu}$, equation (30)) or possibly humidification ($\dot{Q}_{hum}$), offset indoor heat losses ($\dot{Q}_{env}$) and compensate the sensible heat loss through air renewal ($\dot{Q}_{a\_renew}$, equation (29)).

$$\dot{Q}_{a\_renew} = \tau_{fa} \dot{m}_{a,Tot} C_{pa} (T_a - T_{ext}) \tag{29}$$

$$\dot{Q}_{dehu} = \dot{m}_{H_2O} L_v \tag{30}$$

Firstly, the influence of the dewpoint temperature on the total energy demand is presented. Assuming constant outdoor conditions that are representative of a dry winter for a Mediterranean climate, the energy demand is computed for four values of the dewpoint temperature. These values represent both indoor humidity and the thermal insulation level of the envelope. The two values are therefore plotted on the x-axis in Figure 9, which illustrates that a poor building envelope results in a lower dewpoint temperature and consequently in a higher temperature difference with the water temperature setpoint. The air renewal flow rate is computed with equation (26) assuming the absence of humidification / dehumidification demand.

A first predictable result is that a poor envelope results in higher energy demand for air heating, because the heat losses through the envelope are higher and must be balanced in order to satisfy the indoor temperature air requirement. A secondary outcome, perhaps less obvious, is that the energy balance for water is also affected. Poorer envelopes result in lower indoor wall surface temperatures, which necessitates a reduction in the indoor humidity ratio to prevent condensation. As demonstrated in section (2.2), $\omega_{sw} - \omega_a$ is the primary cause of evaporation. Consequently, there has been an increase in evaporation losses in water, as well as an increase in energy demand to maintain its temperature constant. At the same time, the air renewal has to be increased in order to compensate for increased evaporation, leading also to an increased energy demand for air.

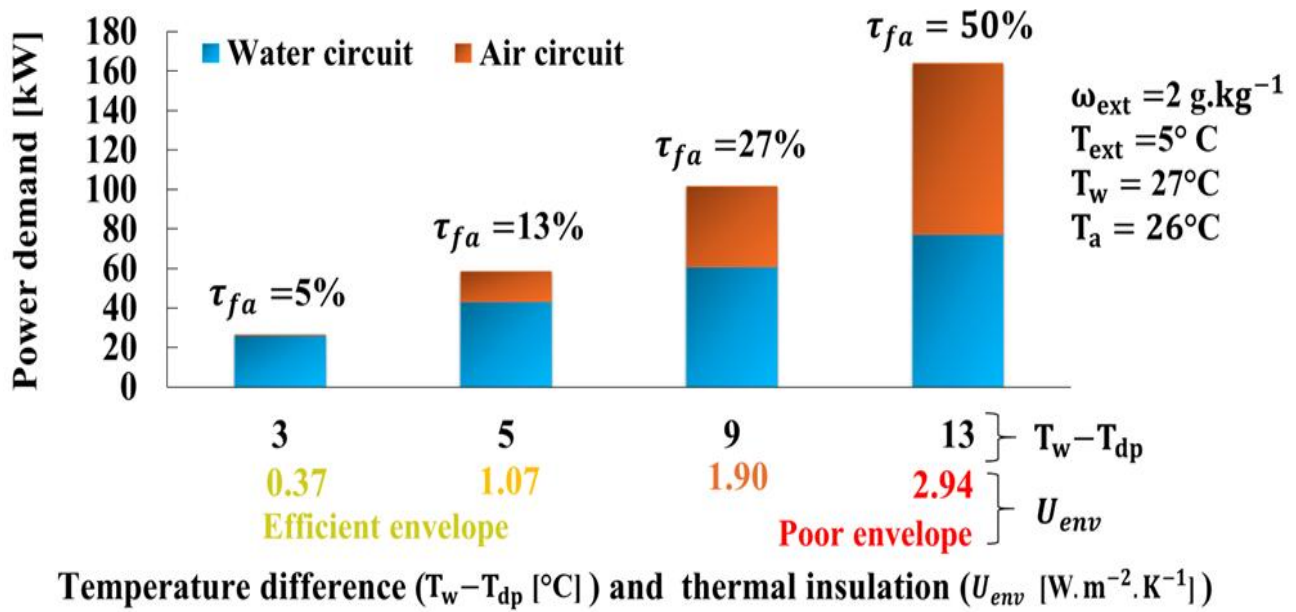

**Figure 9: Total pool power demand as a function of $T_w$-$T_{dp}$ or $U_{env}$**

Next, the value of $U_{env}$ was fixed to $0.81\ W.m^{-2}K^{-1}$ and the fresh air ratio was maintained to 30%, in order to meet the actual values of the case study. The calculations were repeated for the outdoor conditions presented in Table 2 in order to analyze the energy demand. The latter is finally presented in Figure 10.

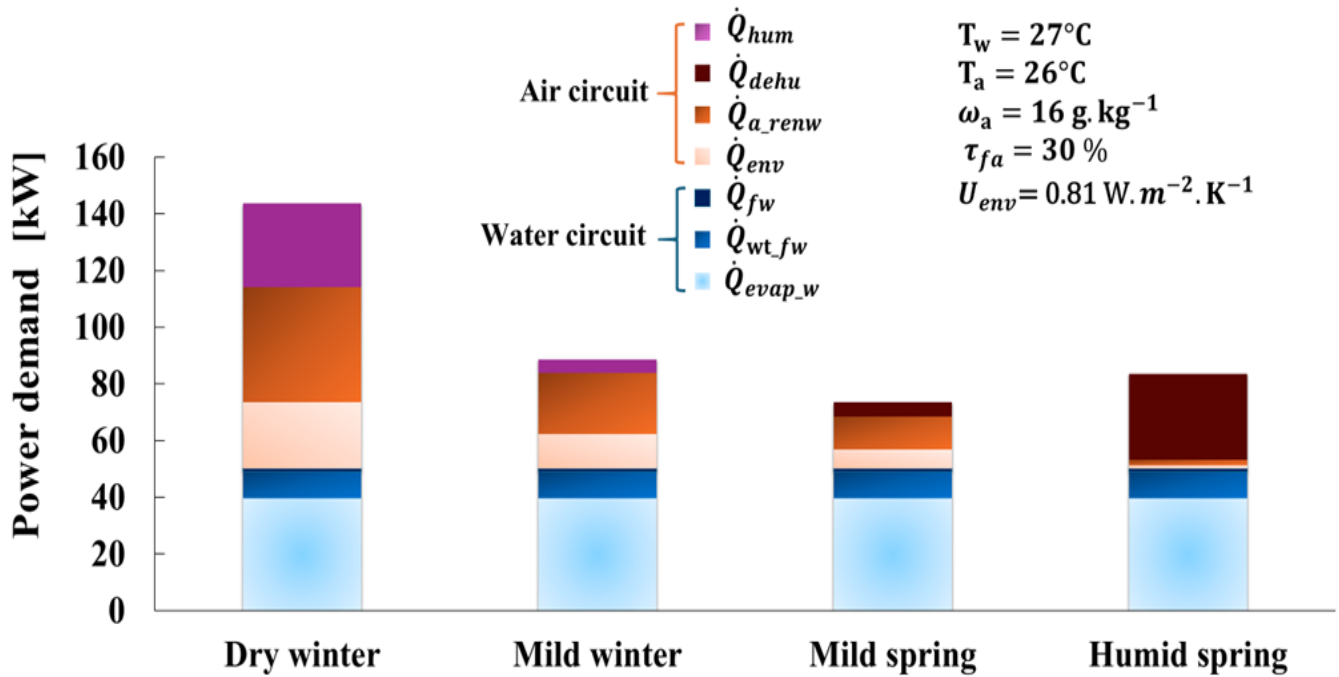

**Figure 10: Power demand for four outdoor conditions**

An examination of the global energy demand reveals that the impact of external conditions aligns with anticipated patterns. Specifically, milder weather conditions are associated with a decline in energy demand. However, an analysis of the decomposition of the energy balance reveals a more mitigated influence.

Firstly, the energy demand for water heating is found to be almost entirely unaffected, a phenomenon that can be explained by the fact that the predominant mechanism for water cooling is related to the indoor conditions, which remain constant. As for air energy balance, the elements related to sensible heat (envelope heat losses, heat losses through air renewal) also decrease for milder outdoor conditions, but this decrease is partially offset by an increase in the energy demand for dehumidification. In fact, the rate of evaporation remains constant throughout the year, but as outdoor conditions become milder, the removal of moisture by air renewal becomes less effective, leading to an increased demand for dehumidification. As already observed, one potential solution to this issue is to adapt the air renewal mass flow rate, but another option is to rely on an energy recovery technique based on dehumidification. This approach is supported by the fact that dehumidification is often achieved by cooling the air below the point of condensation, which requires the removal of heat. The implementation of this process can be advantageous in scenarios where there is a significant demand for heat simultaneously, such as in the SP. Ultimately, the dehumidification rate could be equal to the evaporation rate, thereby allowing the energy expended in condensation to be utilized for reheating the pool and compensating the latent heat associated with evaporation. This is illustrated in Figure 10 for mild outdoor conditions (humid spring), where the energy demand for dehumidification nears the heat loss caused by evaporation. Also, Figure 10 shows that there is a humidification demand ($\dot{Q}_{hum}$) in winter due to the fraction of fresh air chosen by the pool manager (30%), which allows more dry air to enter the pool interior. To avoid humidification/dehumidification in winter, a fraction of 14% could be applied, as determined in Figure 8, but this does not eliminate the risk of $NCl_3$. Increasing the fresh air fraction from 14% to 30% leads to a humidification demand, which is the price to pay for better indoor air quality.

The third and final analysis proposed in this section involves a comparison of two sets of setpoints under two distinct outdoor conditions, while all other parameters remain constant. The setpoints that have been outlined so far will be designated as "Case Study". They were then compared to setpoints that had been more commonly reported in the scientific literature, such as those given in (Delgado Marín et al., 2019), namely 28 °$C$ for water temperature, 30 °$C$ for air temperature and 15 $g_v . kg_{da}^{-1}$ for indoor humidity. This second configuration is denoted as "Literature" in Figure 11.

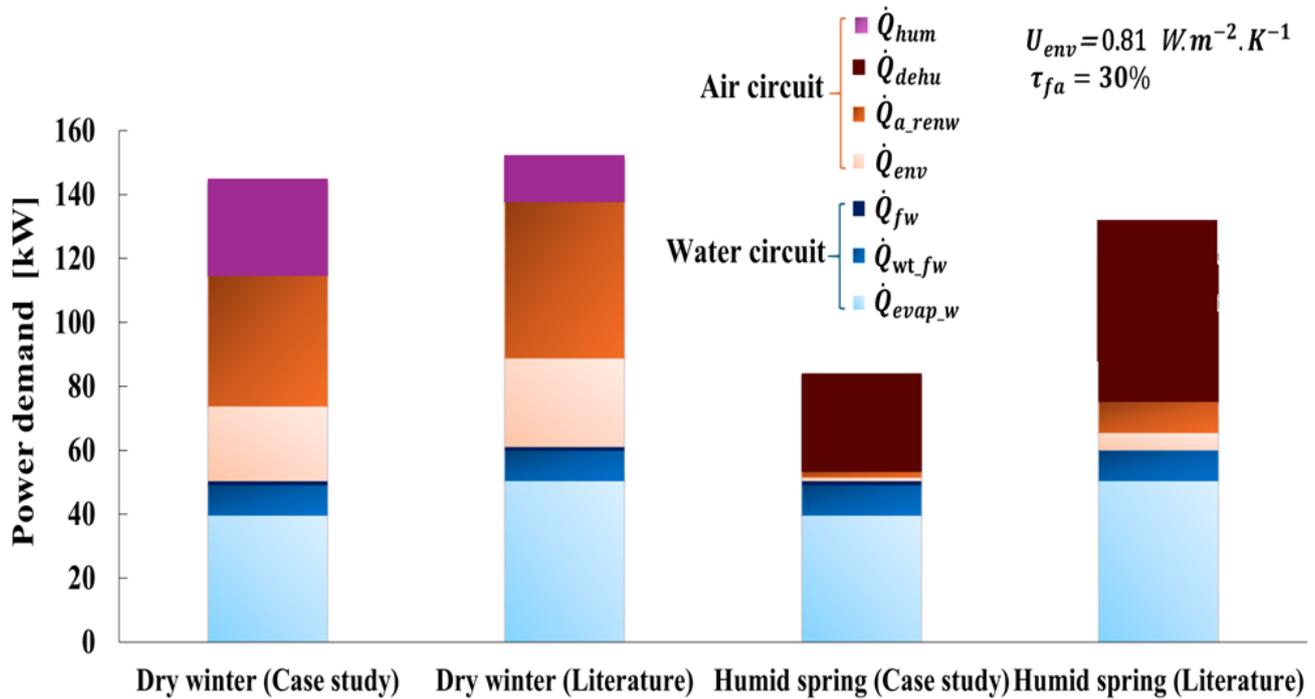

**Figure 11: Power demand depending on the selected setpoints for two outdoor conditions.**

Despite the relatively small variations in the selection of the setpoints, the total power demand varies as much as 5% for winter conditions and 47% for mild conditions. The water energy balance remains unaffected by the outdoor conditions, as already observed, and the power demand increases by 22% when the setpoints are varied because of an increased evaporation rate. In contrast, the air energy balance is found to be more significantly impacted, exhibiting an increase of 84% under milder outdoor conditions and a decrease of around 4 % under harsh conditions. This small decrease results from the higher humidification demand for the case study, because of the lower evaporation rate. However, it is clear that despite the same thermal insulation level, there is a significant increase in power demand in winter due to the need for heating the fresh air introduced into the building.

This analysis underscores the effectiveness of the strategy advocated by the pool's manager, which aims to achieve energy savings by means of reducing the indoor air and water setpoints. The conclusions of the present chapter can be summarized as follows:

- Evaporation has two consequences for energy demand: firstly, the pool must be heated to compensate for latent heat loss; secondly, the indoor air must be dehumidified to remove

- excess moisture caused by evaporation. This is especially true under mild outdoor conditions, when air renewal performs less efficiently;
- The quality of the thermal insulation plays a dual role. Firstly, it reduces energy losses through the envelope, and secondly, it increases the indoor surface temperature as well as the indoor dewpoint temperature. Consequently, the indoor humidity can be set at a higher level, thereby enabling the reduction of energy losses caused by evaporation. Therefore, enhancing the envelope performance appears to be a key aspect in the design and refurbishment of ISPs;
- It is evident that the water temperature setpoint plays a major role in the global energy demand, as it also drives evaporation. A reduction of merely 1°C in water setpoint temperature can result in a significant energy demand reduction. Nevertheless, the impact on the comfort of swimmers may also prove to be considerable, although this issue falls outside the scope of the present study.

This analysis provides a clear understanding of the main phenomena that affect the energy demand of a SP. However, it is limited to a steady state and under a few outdoor conditions. To address this shortcoming, a numerical model has been developed to calculate the energy demand under dynamic conditions for a whole year and to allow a more in-depth analysis.

## 5 Numerical model for a dynamic and time dependent analysis

The dynamic thermal simulation software TRNSYS (Beckman et al., 1994) was selected for the modelling of the SP following a benchmarking process involving several software tools, including a comparison with Dymola (Benakcha et al., 2024). It was determined that both software packages had the capability to simulate a complete SP, including its equipment and associated control systems. The selection of TRNSYS was primarily influenced by its extensive HVAC library and, most significantly, its rapid computation speed.

The numerical model for the water loop was developed in accordance with the diagram presented in Figure 5. The TRNSYS diagram is presented in the appendices section 10 and contains the following elements:

- The solar loop comprises a pump (Type 114) in conjunction with solar collectors (PVT, Type 50b), with a water glycol mixture serving as the heat transfer medium. The total surface area of the solar collectors is divided into two groups. The inclination of the panels relative to the horizontal surface of the roof is constant (10°) for both groups, while their orientations are opposite. One half has an orientation of 36°, and the second half has an orientation of −144° relative to south (0°). This fixed configuration of the panels was chosen and imposed to respect the existing architectural and structural constraints of the roof as well as for optimal solar performance. Consequently, the objective of the study is to analyse the system under these real-world constraints.
- The pool water loop is composed of: a single pump for the main water circuit (Type 114) with a mass flow rate of $191 \ m^3 h^{-1}$; two Type 158 tanks to model the pool water volume and the buffer tank; a pair of valves to model the water flow lost through evaporation and renewal (Type 11f) as well as its compensation (Type 11h) with fresh water from the city network at the temperature of the ground. Finally, a heat exchanger (Type 91) that connects the main circuit with the solar circuit. A set of two valves (Type11f + Type11h) was used to deviate a fraction of the main circuit to the heat exchanger connected to the solar circuit.
- The schedule of occupation is modelled on the actual schedule of the ISAE SP, as shown in Figure 12. The evaporation rate depicted in equations (12) and (13) has been implemented in TRNSYS using logical programming, and uses the schedule of occupation accordingly.

- Finally, a variable ground temperature was considered using Type 77.

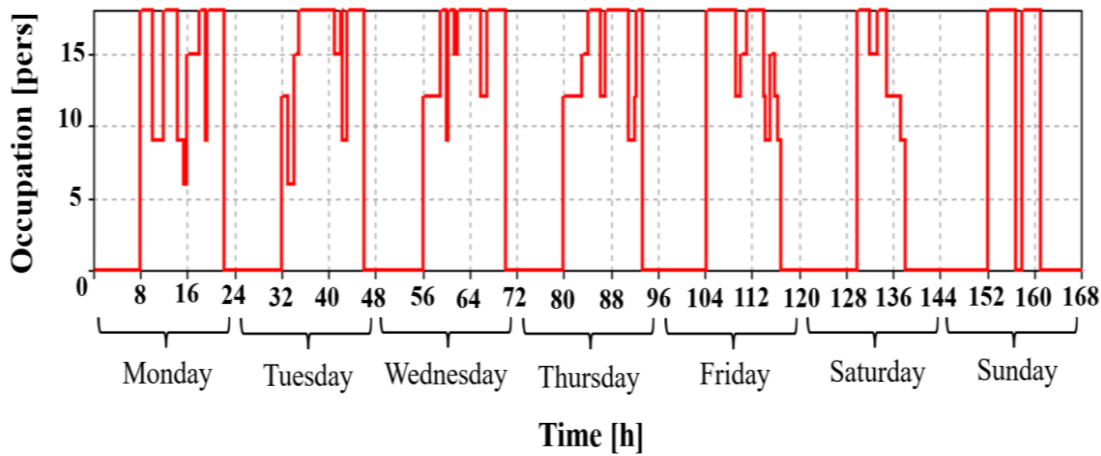

**Figure 12: Weekly occupancy of the ISAE SP**

The pump for the solar loop was activated when solar radiation exceeded $100\ W.m^{-2}$. The distribution valve was switched on when there was a demand for heating and when the temperature in the solar circuit exceeds the temperature in the SP by at least $15°C$. It should be noted that several amendments were made to facilitate the modelling of heat transfer in TRNSYS:

- In order to consider the thermal inertia of the fluid volume in the pipes, two small tanks were added to the solar loop (Type 158) in TRNSYS.

- The pump for diverting the flow fraction toward the heat exchanger (also called heating circuit) was not modelled, as the use of two valves has been demonstrated to result in the same outcome in TRNSYS.

- The district hot water network was modeled by an idealized heat source, namely a boiler without heat losses (Type 659), with a maximum heating power of 220 kW. The temperature of the pool was maintained at the setpoint value by a PID controller (Type 23) throughout the entire year.

The simulation was conducted over the entire year for the temperate climate of Toulouse, which has both oceanic and Mediterranean influences. The computed energy consumption by the water circuit and its distribution on different posts is illustrated in Figure 13.

It was determined that the auxiliary heat source for heating water was the most energy-consuming, accounting for approximately 55% of the total energy demand in the water loop. A proportion of this demand was met by solar panels, and to a lesser extent by the metabolism of the swimmers as shown in Figure 14.

Filtration pumps are the second most energy-consuming equipment, accounting for 40% of the total demand. Whilst this figure may appear high, it is in line with the results provided by (Yuan et al., 2021) and (Nikolic et al., 2021), who reported 25% and 30% respectively. In their study, however, domestic hot water and lighting were included in the total energy demand. Besides, it should be noted that the filtration pumps operate continuously at nearly half of their capacity, indicating that this issue has already been identified as critical for energy consumption. It is reasonable to conclude that reducing the mass flow rate further would lead to additional energy savings but would have a negative impact on water quality.

The heating circuit pump operates continuously at its nominal power. Although it is not physically modeled as a component in TRNSYS, its consumption is accounted for through a separate calculation within TRNSYS. It represents approximately 4 % of the total energy consumption was the energy consumption for the solar pump was estimated to be negligible (less than 1 %). This study did not consider additional electrical consumption units, such as lighting, electrical plugs and domestic hot water pumps, because their consumption is not significant and does not affect the coupling of physical phenomena between water and air.

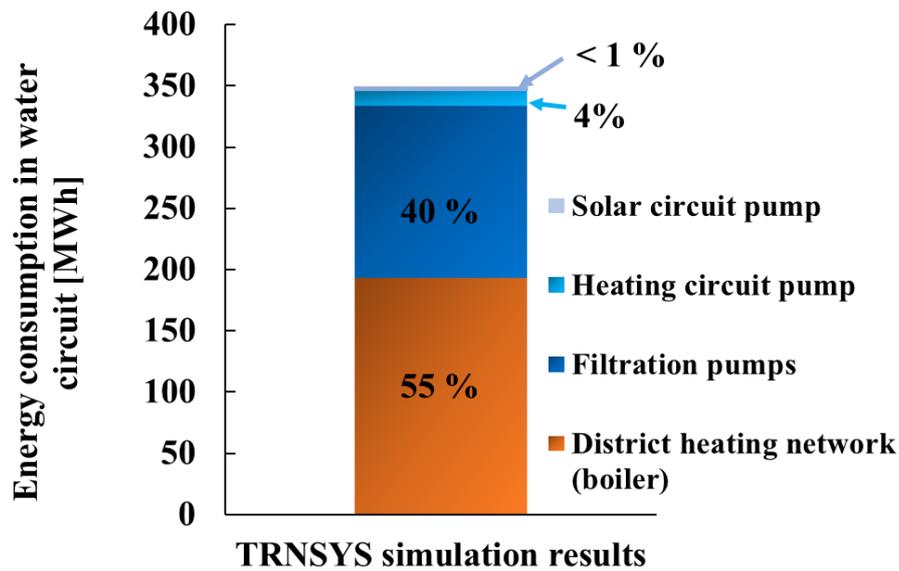

**Figure 13: Energy demand for water for a whole year**

Next, the water energy balance is presented in Figure 14, based on the equations presented above. For the sake of clarity, heat gains and losses have been segregated into two distinct graphics. With regard to heat losses, evaporative losses account for 67%, a finding that is consistent with the results reported in the scientific literature, especially in (Ribeiro et al., 2016) and in (Nikolic et al., 2021). The second major component of heat loss is attributed to water renewal, accounting for 19% of the total heat loss. As previously observed, these figures may vary significantly depending on national standards, with water quality being a primary contributing factor. In this study, the air temperature setpoint is very close to that of the water and results in low convection and radiation heat losses (5%). The consequence is that more substantial conductive heat losses (7%) are observed. Finally, the process of replenishing the evaporated water volume with fresh water results in a minor heat loss of less than 2%.

In terms of heat gains for the water balance, the largest heat source is the additional thermal energy provided by the district heating (boiler), which contributes 65% of the total annual heat gain. This share is relatively high, but it could be lower in other pools that use more than two heat sources for water heating. This is exemplified by the ISAE pool, which employs a third energy source in the form of a heat recovery unit integrated with dehumidification technology. Latent heat is transferred back to the pool water through a heat exchanger. This source is not considered in the present study and is instead accounted for as part of the boiler's contribution. This issue will be addressed in a forthcoming paper focusing on the air circuit, and forms part of the overall perspective of this work.

A deeper analysis of the simulation results reveals that the boiler is used at between 12% and 20% of its maximum capacity ($220\ kW$) during cold periods, particularly in winter. The availability of the boiler's power makes it suitable to implement a restart strategy for the boiler, following a night-time shutdown. This allows the water temperature setpoint to be reached quickly when occupants arrive at the pool in the morning. This approach will be evaluated in Section 6.

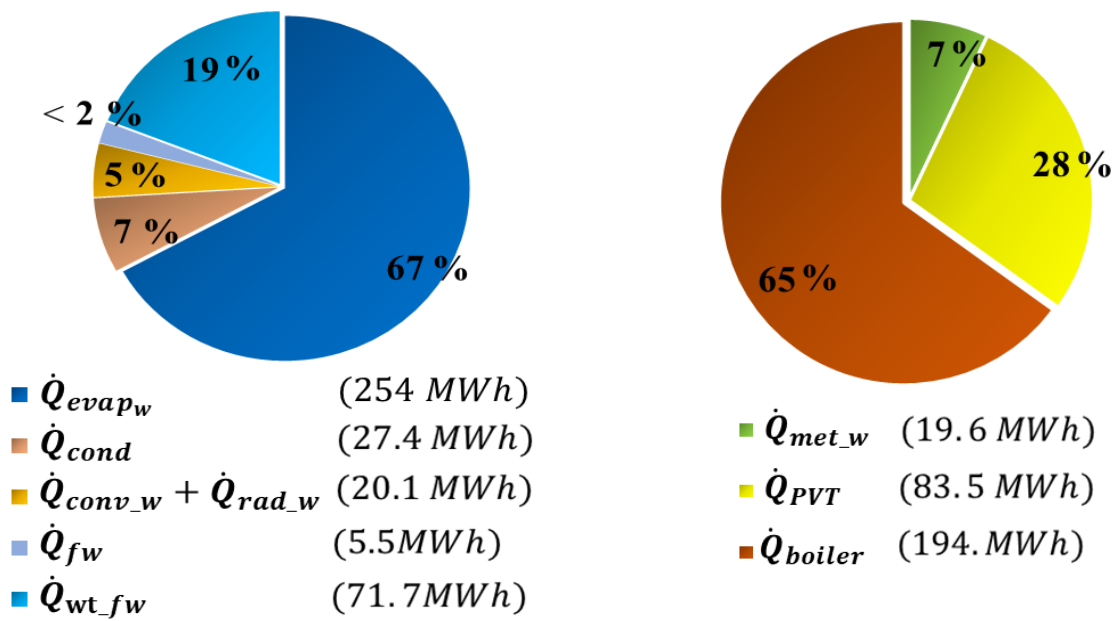

**Figure 14: Heat losses (left) and heat gains (right) in water for a whole year**

The heat gains attributed to the swimmers account for 7% of the total heat gains. This value may be slightly overestimated, as it was assumed that all swimmers develop an intense muscular effort, which may not be the case. However, even if this value were halved, it would not be negligible.

With regard to solar panels, they provide 28% of the heating demand for water, which is approximately 83 $MWh$. However, the available solar resource accounts for as much as 682 $MWh$, meaning that only 12% was recovered as heat. This figure is relatively low, indicating that solar energy is not being utilized to its full potential. As previously outlined, PVT technology was employed, with a nominal electrical efficiency of 20%, leading to an estimated yearly production of 89 $MWh$ of electricity and a global energy recovery of 25%. In the actual building, electricity is shared on the grid, but it should be noted that it accounts for almost 63% of the energy consumption of the filtration pumps. However, the heat recovery from the PVT remains modest. While heat losses to the environment cannot be avoided, the main reason for this is the mismatch between the heat demand and the solar resource, which is a well-known issue when it comes to renewable energy. However, the large volume of water in the pool could be utilized as a means of storing heat, provided that temperature variations can be accommodated. This will be studied in greater detail in the next section.

## 6  Potential for energy savings

In this section, the potential energy gains for two different strategies will be numerically assessed. One strategy is to explore the advantage of maximizing the water temperature under high solar heat loads using solar panels. As previously demonstrated, the PVT surface is large enough to compensate for a significant fraction of the heat demand, but a high fraction of the solar load is not recovered because the water temperature has already reached the set point. Conversely, when solar loads are reduced, the PVT system is unable to meet the heat demand, necessitating the provision of heat by the district heating. The SP can be regarded as a large sensible thermal storage tank due to its significant mass and the water's high specific heat capacity (about $4.18\ kJ \cdot kg^{-1} \cdot K^{-1}$). In this case study, an increase in temperature of merely $1°C$ would necessitate the consumption of approximately $700\ kWh$ of energy, because of the substantial water volume of $581\ m^3$. Consequently, even minor temperature fluctuations in the pool can result in substantial energy fluctuations, which might impact significantly the overall heating demand. It was therefore foreseen that allowing overheating by means of the solar circuit would reduce the energy supplied by the district heating network. This strategy will be referred to as "overheating" later in this paper.

The other strategy that is examined in this paper is analogous to the one implemented in common buildings, where a night setback temperature setpoint is employed during periods of unoccupancy. As the indoor temperature declines, so do the heat losses, thereby enabling energy savings. However, it is necessary to manage the set-back so that the indoor temperature reaches the desired value at the start of occupancy. A similar strategy was tested here, whereby the water temperature was permitted to decrease below the setpoint during periods of unoccupancy. This second strategy is referred to as "subcooling" in this paper and is depicted in Figure 15. In (Delgado Marín et al., 2019), a more complex approach, derived from MPC (Model Predictive Control), was used without quantifying the effect of overheating, for different water temperatures, on minimizing heat demand. Also, the authors focused on boiler shutdown time. In contrast, the method presented in the present paper determines the optimal boiler restart time which ensures that water reaches the setpoint temperature at the beginning of occupancy, using a simpler approach. Also, the present paper assesses the effect of water overheating at different temperatures on energy consumption, as well as the combination of the two strategies (Sub-cooling and Overheating). Rev. 3 Com .2

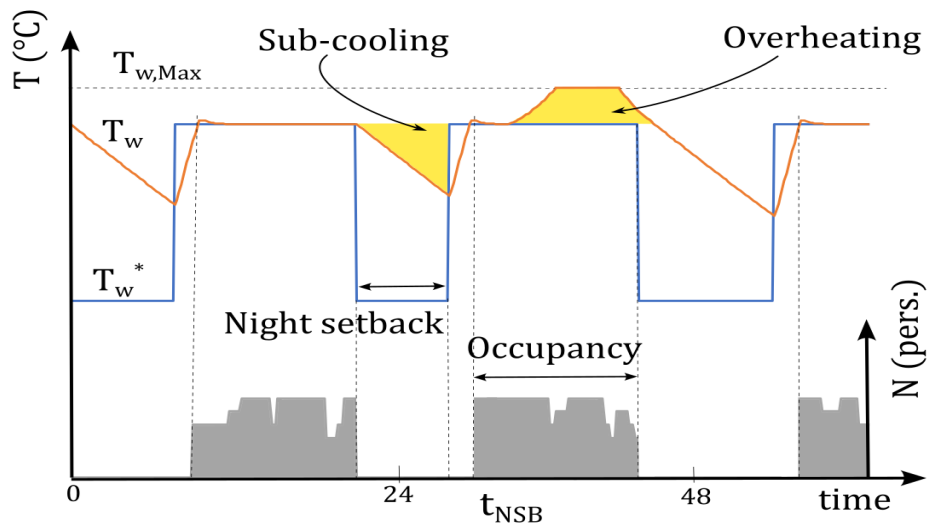

**Figure 15: Energy saving strategies that enable overheating during periods of occupancy and subcooling during periods of unoccupancy.**

For the sake of clarity, the method used to adapt the TRNSYS model is not presented here and was moved in section 10.

The results are analyzed by means of two criteria. Firstly, the time of end of the night setback ($t_{NSB}$) was used to plot the histogram in Figure 16. Secondly, the annual energy consumption of the auxiliary heat source was computed for each strategy (5 overheating temperatures, with and without a night setback) leading to 10 combinations. The relative energy demands are then compared in Table 3. The reference case is the same as presented in Section 5 (i.e., with no overheating and a constant water temperature setpoint).

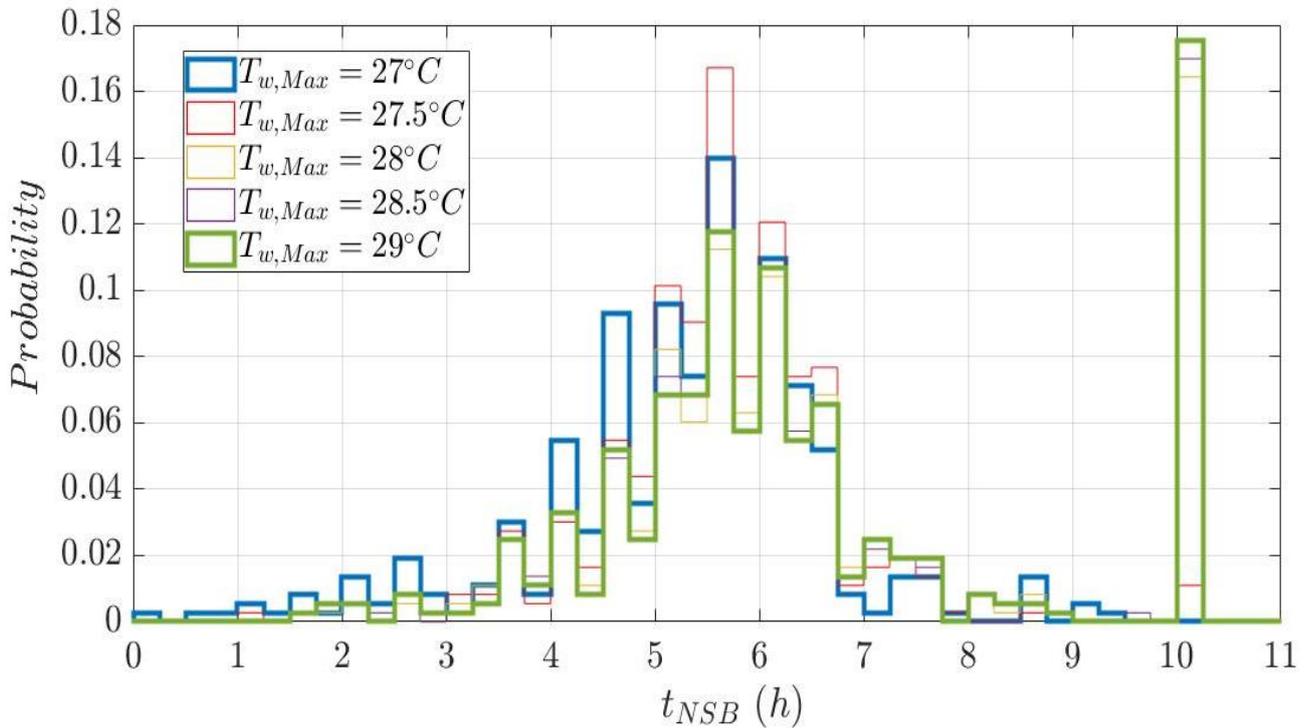

**Figure 16: Distribution of $t_{NSB}$ over one year and for different values of $T_{w,Max}$.**

With reference to $t_{NSB}$, when overheating is not permitted ($T_{w,Max} = 27\ °C$), the distribution appears to be scattered, with peak values occurring between 5 and 7 a.m. This is consistent with the minimal pool temperature observed during the night (approximately $26.35°C \pm 0.1°C$ during the year) and the maximum heating power of the auxiliary source ($220\ kW$). Utilizing the latter at maximum capacity would require slightly over two hours to elevate the pool temperature to $27°C$ at the onset of occupancy (8 a.m.). The optimized values for $t_{NSB}$ exhibit greater contrast, however, for the following reasons:

- The heat losses vary over the year (the ground temperature and the water temperature for refilling);
- The duration of unoccupancy varies during the week, and occupancy starts later on Saturday;
- During summer, the solar loop might provide some heating before occupancy, delaying $t_{NSB}$;
- The use of a PID controller smooths the heat demand when the temperature setpoint is approached, thereby necessitating earlier initiation of heating.

When overheating is allowed, $t_{NSB}$ is generally shifted towards higher values, which means that heating can be started later. This becomes particularly apparent when the maximum temperature is 28°C or higher, as the maximum value for $t_{NSB}$ (10 a.m.) is often selected: up to 17% of the year. This phenomenon occurs after occupancy begins, regardless of the day of the week, which means that there is no need for heating during the night. This phenomenon is in line with the figures presented above, since a temperature drop of less than $1°C$ was observed during unoccupancy. Consequently, if the pool is heated up to $28°C$ the day before, the energy stored within the pool water is sufficient to compensate for the heat losses during unoccupancy. It is also observed that increasing the maximum temperature from 28 to $29°C$ does not lead to significant changes, although some small improvements can be observed.

**Table 3: Relative heat demand when the high temperature threshold for water varies and for constant or variable water temperature setpoint.**

| $T_{w,Max}$ (°C) | 27 | 27.5 | 28 | 28.5 | 29 |
|---|---|---|---|---|---|
| $T_w^* = c^{te}$ | 1⋆ | 0.880 | 0.827 | 0.820 | 0.817 |
| $T_w^*$ = variable | 0.957 | 0.848 | 0.807 | 0.800 | 0.797 |
| ⋆ Reference case | | | | | |

The heat demand is now examined through the values presented in Table 3. Compared to a constant water temperature setpoint, a significant drop in the heating demand (12%) is observed when $T_{w,Max}$ is increased from 27 to 27.5°C. It indicates that solar energy is favored when overheating is permitted. A further decrease in heating demand was observed for a maximal threshold of $28°C$, reaching 17% reduction. However, it was noted that exceeding this threshold did not result in significant enhancements. This finding aligns with the results presented above, which demonstrate that when the pool temperature reaches $28°C$, the energy stored is sufficient to compensate for the heat losses during the night, thereby eliminating the need for additional heat sources until the next day when solar energy becomes available again. Overheating above 28°C is helpful only when the next day is cloudy, without sufficient solar gains to compensate for the heat losses. Although overheating enables achieving significant energy savings for the pool water circuit, it should not be forgotten that it also increases evaporation. Overheating the water to $28\ °C$ significantly reduces the energy demand. Increasing it further becomes quickly less beneficial, which may physically justify the frequent recommendation of this value in the literature. In fact, increasing water temperature enhances buoyancy-driven natural convection near the water surface. As the air just above the surface becomes saturated and less dense, it rises, while drier, denser air descends to replace it, thereby accelerating evaporation. This leads to a greater need for dehumidification, ventilation, and potentially cooling in the pool's air circuit. These effects have not been studied in this work and will be the subject of a detailed analysis in a second article on the air circuit. This might give the impression that the energy gains achieved on the pool's water circuit are offset by dehumidification heat demand on the air circuit, which is not always true. In fact, there are ways to recover the energy released during dehumidification in some air handling units, as is the case with the system used in the ISAE pool.

Second, it was observed that the consideration of a variable water temperature setpoint (with the corresponding value for $t_{NSB}$), depending on overheating temperature resulted in energy savings ranging from 2 to 4%, depending on the value employed for $T_{w,Max}$. While energy savings may be more modest with subcooling, the economic benefits could still be substantial, particularly in the light of the significant heat demand for water. It is interesting to note that the strategy allowing overheating would be very easy to implement and could lead to significant energy savings as soon as $1°C$ of overheating would be allowed. In contrast, the subcooling strategy is more complex to implement, as it necessitates precise calibration of the $t_{NSB}$ to optimize energy savings while maintaining strict comfort conditions. The combination of both strategies, with an overheating temperature of $28\ °C$, can result in energy savings of approximately 19%, as can be determined from Table 3.

# 7 Discussion

To achieve energy savings in SPs particularly and in buildings more generally, other researchers use more advanced control and optimization techniques than the equipment management strategy presented here. Trend control and optimization methods in scientific literature are Neural Networks (Yuce et al., 2014), Particle Swarm Optimization (Lee and Kung, 2008), Fuzzy Logic, Genetic Algorithms, and Reinforcement Learning (Wang and Ma, 2008). The challenge with these techniques lies in their status as a research area, which makes them difficult to apply in practical contexts. These methods are rarely implemented in real control and automation systems because they are resource-intensive and require significant computational capacity and modeling effort, given that several input and output data in the pool must be processed. The author's work, through this first article, is precisely aimed at identifying methods that can be easily applied in the field and at replacing commands derived from complex modeling with simple setpoints that can be implemented in actual pool automation systems. Therefore, the authors believe that the findings, particularly the simple strategy of allowing overheating, can be generalized to other indoor pools with different dimensions, climatic conditions, or energy systems, provided the pool is equipped with a solar heating installation. Especially, when it doesn't require replacing the existing equipment with more efficient alternatives. Nevertheless, further investigations are required to confirm whether the repercussions of overheating on the air circuit can be managed for a low energy cost. This falls within the perspectives of this work and the future scope of a second article. The latter will, this time, include preferably simple strategies to reduce energy consumption in the air circuit. It will also feature experimental validation carried out on electrical and thermal consumption data collected from the pool manager, as well as temperature and humidity data measured with external KIMO data loggers. This validation could not be included in the present article because it involves a coupled air-water circuit with a variable air temperature, which was not studied here.

## 8 Conclusions

Through a case study, this paper investigated the energy dynamics of ISPs, emphasizing evaporation, insulation, and setpoint control as key factors affecting energy demand. Firstly, through a steady-state study it was demonstrated that improving thermal insulation not only reduces heat loss but also allows higher indoor humidity, thus mitigating evaporation losses. Water temperature setpoint significantly influences total energy use, with even small reductions. Decreasing it from $27°C$ to $26°C$ resulted in a 26% decrease in evaporation rate and yielding substantial savings, under the assumption of a constant dew point temperature.

Secondly, dynamic simulations have shown that district heating supplies most of the water's thermal energy, roughly two thirds, though solar panels also play major roles. Filtration pumps are one of the highest energy-consuming components in a pool. Evaporative losses dominate heat demand, accounting for two-thirds of water heat loss, followed by water renewal. Solar contributions are underused due to fixed temperature limits mostly.

Two strategies were proposed and tested to increase the heat recovery of solar panels: overheating during periods of occupancy and subcooling during periods of unoccupancy. Allowing slight overheating ($1°C$) enables significant heat gains (17% throughout the year) while subcooling reduces night-time losses, but to a lesser extent with maximum of 4%. This last value seems low compared to the findings (19%) in (Delgado Marín et al., 2019), but this can be explained by the fact that the control methods are not exactly the same and the two studied pools are different, especially in terms of external climatic conditions, occupancy, pool tightness, etc. Finally, the authors subscribe to the view that the strategy of controlled overheating can be extended to other ISPs with solar heating systems, regardless of size or climate, due to its ease of implementation.

# 10  Appendices

## 10.1  Water circuit model in TRNSYS

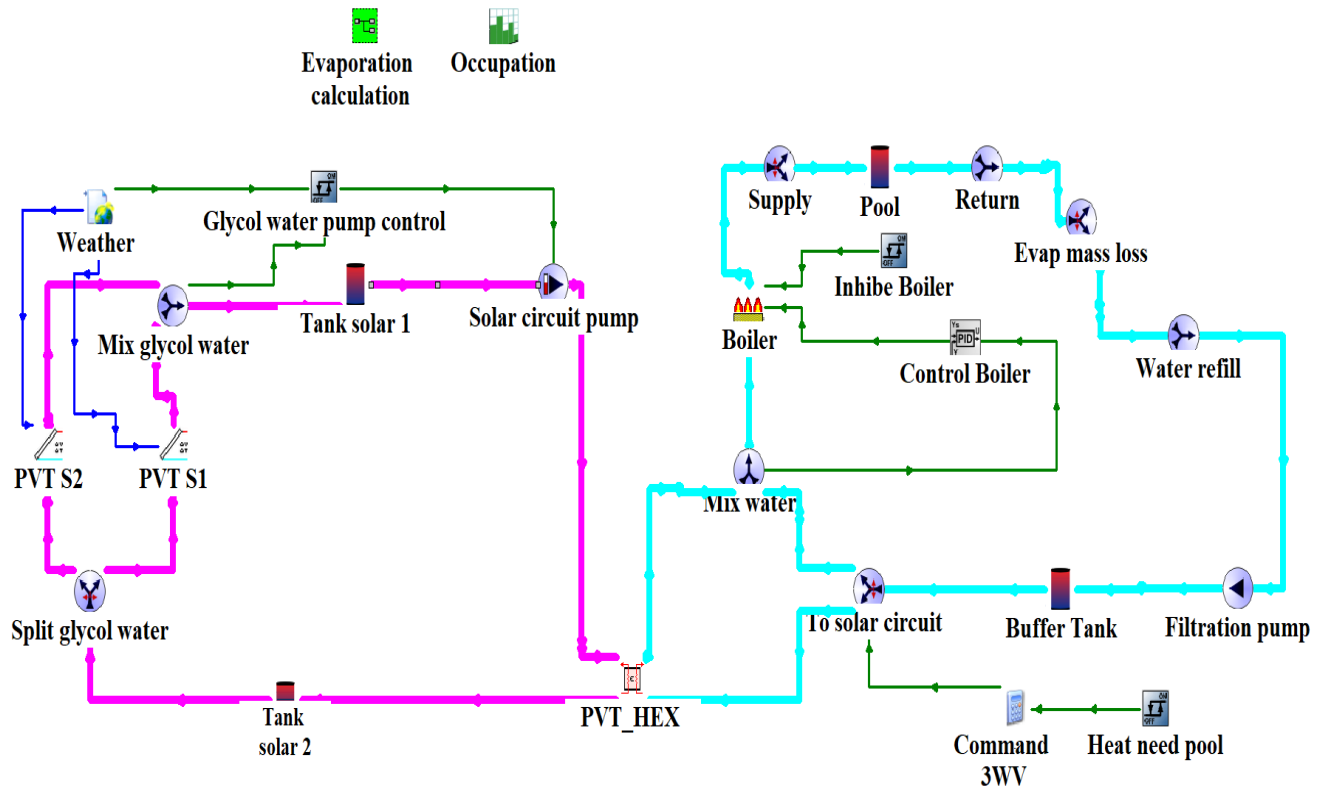

## 10.2  Method used to model energy saving strategies

During the preliminary testing phase, it was observed that the total energy consumption exhibited a high degree of sensitivity to the configuration of the PID controller and the control of the diverter when overheating was not permitted. Variations in energy consumption were recorded as high as 10%. This phenomenon can be attributed to the fact that two different controllers were implemented: PI controller for the auxiliary heat source, which adapts its power output when the setpoint is almost reached, and a differential controlled with hysteresis for the solar loop. Consequently, when solar energy is available and being transferred to the primary water loop, the pool water setpoint can be rapidly exceeded, leading to the solar loop being bypassed for the subsequent time step (100 seconds in this particular study). However, solar energy should be prioritized. To address this issue the PI controller was paused when the water temperature exceeded $27.05°C$, which is lower than the observed overshooting for step variations and with no solar energy available. Second, the high limit cut-off for the diverter of the solar circuit was set to at least $27.1°C$. This minor modification was found to be advantageous for the utilization of the solar loop, resulting in a significant reduction in the occurrence of the aforementioned issue.

The precise management of the night setback temperature setpoint is challenging. Due to the varying occupancy and outdoor conditions, it is preferable to identify a variable hour for heating restart, rather than defining a fixed restart time. In this paper, it was proposed to breakdown the annual problem into 24-hour simulations. This approach builds upon earlier research conducted for a different system with high thermal inertia, namely a ventilated slab (Nevers et al., 2025), wherein the MATLAB, TRNSYS, and JE+ software were combined to execute multiple simulations concurrently and identify the most suitable option. Particular attention must be paid to initial conditions when starting a new 24-hour sequence, which should be equal to the final state of the previous sequence. This has necessitated the modification of certain TRNSYS libraries written in Fortran, such as the water tank and PID libraries. These modified components have been validated by comparing the results of a single 1-year simulation with 365 24-hour sequences. It was found that the root mean square error for the temperature of the pool between the two cases over a year-long simulation was $1.10^{-3}\ °C$, with the maximum error remaining below $3.10^{-2}\ °C$, indicating that the approach is reliable.

The night setback temperature was set at $25°C$, as the time required for the water temperature to increase by $2°C$ is approximately 24 hours, which exceeds the maximum time for inoccupancy. This ensures that there will be no heat demand during night setback. In order to define the optimal night setback duration, a systematic approach was adopted, where the time for resetting the temperature setpoint ($T_w^*$) was tested for every 10 minutes and for every day. This approach consequently resulted in the execution of a simulation exceeding 22,000 instances, which took about 4 hours of computation time, thanks to the numerical model and the techniques used. The identification of the optimal setback duration for each 24-hour sequence was achieved through the selection of the solution that exhibited the minimum energy consumption within that sequence. This was done while ensuring that the water temperature setpoint was met at the beginning of the occupancy period.

Finally, these two strategies were combined, and this computational work was repeated for five different values of the maximum temperature allowed for the pool ($T_{w,Max}$), ranging from 27 to 29 °C by steps of 0.5 °C. For each value of $T_{w,Max}$, the restart time was tested for 24-hour sequence as explained above.